\documentclass[superscriptaddress,twocolumn,reprint,amsmath,amssymb,aps]{revtex4-2}
\usepackage{graphicx,physics,xcolor}

\newcommand{\h}[1]{\hat{#1}}
\newcommand{\expv}[1]{\langle #1 \rangle}

\begin{document}

\title{Nonlinear Dynamics of a Dicke Model for V-Type Atoms}

\author{Ofri Adiv}
\affiliation{Department of Physics, University of Auckland, Auckland 1010, New Zealand}
\affiliation{Department of Mathematics, University of Auckland, Auckland 1010, New Zealand}
\affiliation{Dodd-Walls Centre for Photonic and Quantum Technologies, New Zealand}

\author{Bernd Krauskopf}
\affiliation{Department of Mathematics, University of Auckland, Auckland 1010, New Zealand}
\affiliation{Dodd-Walls Centre for Photonic and Quantum Technologies, New Zealand}

\author{Scott Parkins}
\affiliation{Department of Physics, University of Auckland, Auckland 1010, New Zealand}
\affiliation{Dodd-Walls Centre for Photonic and Quantum Technologies, New Zealand}

\date{\today}

%###########################################
%                Abstract
%###########################################

\begin{abstract}
We study the nonlinear, semiclassical dynamics of an open spin-1 (three-level) variant of the traditional Dicke model. In particular, we focus on V-type energy-level configurations with varying degrees of energy-level asymmetry. We also allow for unbalanced coupling -- where co-rotating and counter-rotating Hamiltonian terms are independently tunable. We characterise the system with a dynamical systems approach, where different behaviors map to definite dynamical objects, and phase transitions to bifurcations. We find the emergence of both periodic and two-frequency oscillations, as well as multistability and chaotic dynamics.
\end{abstract}

\maketitle

%###########################################
%                Introduction
%###########################################

\section{Introduction}

Control of interacting quantum systems has been a focus of research for a long time. Aside from being of fundamental interest, it attracted recent attention due to its relevance for future technologies. In particular, many-body light-matter systems, within the framework of cavity quantum electrodynamics (cavity QED), offer enticing platforms for both research and emerging technologies, because advances in atomic manipulation and cavity engineering allow for an exquisite degree of control over system parameters \cite{mivehvar2021}.

One paradigmatic system of cavity QED is the Dicke model \cite{kirton2018}. It describes the collective interaction of an atomic ensemble of two-level emitters with a single mode of radiation in an optical cavity. Famously, the Dicke model undergoes a quantum phase transition at a critical value of the light-matter coupling strength: from the normal phase, with no steady-state cavity mode excitation, to the superradiant phase, with macroscopic cavity mode excitation.

Since its original description by Hepp and Lieb \cite{hepplieb}, various extensions to the Dicke model have been proposed. Pertinent to us is control of \textit{unbalanced} coupling \cite{dimer,zhiqiang17,stitely_bistab,stitely}, where interactions that conserve the total number of excitations (cavity $+$ atomic) are given a different strength to interactions that change the number of excitations by two.

Imbalance between these interactions, dubbed the \textit{co-rotating} and \textit{counter-rotating} interactions, respectively, leads to interesting dynamics. Experimentally, dominance of the counter-rotating interactions has been shown to cause oscillatory superradiance \cite{zhiqiang17}. This has been backed theoretically through simulation \cite{stitely_bistab}, as well as a bifurcation analysis in the semiclassical (mean-field) regime \cite{stitely}. The latter regime corresponds to the limit of many atoms, where the system dynamics can be described by a set of nonlinear ordinary differential equations. Solutions of these equations can undergo qualitative changes as parameters are varied, as described by bifurcation theory \cite{kuznetsov,strogatz,guck,carr}, which manifest themselves as (quantum) phase transitions of the physical system. For example, the aforementioned emergence of oscillations corresponds to a Hopf bifurcation of an equilibrium (steady-state) of the semiclassical equations of motion \cite{stitely}. The power of the bifurcation-oriented approach lies in its ability to follow solutions as parameters are varied, regardless of their stability. In contrast to simulations, it allows one to describe the dynamics in various parameter regimes without relying on stability and a specific choice of initial conditions. 

In this work we apply this approach to a spin-1 variant of the Dicke model, where three magnetic sublevels, subject to both linear and quadratic Zeeman shifts, are available to each atom. This allows one to consider a wider range of level configurations with qualitatively different dynamics \cite{Adiv2024}. Here we focus on a V-shaped energy level configuration, where the sublevel with quantum number $m=0$ has the lowest energy. The choice to consider spin-1 or three-level atoms may seem arbitrary at first, but such atoms have found a number of uses, including electromagnetically induced transparency \cite*{boller1991,fleischhauer2005}, lasing without inversion \cite*{scully1989}, quantum information manipulation \cite*{lopez2023}, and engineered spin squeezing \cite*{masson2017}. Hence, studying this specific variant of the Dicke model could be of use in a number of applied areas.

Works on similar three-level Dicke models exist from as early as 1979 \cite*{sung1979,brandes2011,brandes2017,skulte2021a,skulte2021b,lin2022,fan2023,padilla2023,valencia2023,adiv}, but none use the aforementioned dynamical systems approach. In the spirit of recent studies of the two-level Dicke model \cite{stitely,stitely_bistab}, here we are able to paint a more detailed dynamical picture of the system and, thereby, contribute to the understanding of light-matter interactions as a whole.

We begin, in Section \ref{sec:model}, with a description of the system, its Hamiltonian, and the resulting semiclassical equations. In Section \ref{sec:sym_v}, we study the symmetric V configuration, where the $m=+1$ and $m=-1$ sublevels are degenerate. In this special configuration we find a comparatively simple bifurcation scenario: equilibrium superradiance emerges through a saddle-node or Hopf bifurcation, depending on the choice of parameters. Moreover, a surface in phase space becomes foliated with energy-preserving oscillatory trajectories when the co-rotating and counter-rotating interactions have equal strengths. We attribute this to a balance of energy exchange between the atoms, the cavity, and the environment. Lastly, we present the phase diagram as a function of the co-rotating and counter-rotating interaction strengths. In Section \ref{sec:asym_v} we explore the effect of asymmetry in the level structure. The aforementioned energy balance is broken, and a complicated set of bifurcations involving both equilibria and oscillatory solutions emerges as a result. These lead to stable two-frequency solutions, multistability, and chaotic dynamics, which we map out in a phase diagram.

%###########################################
%                Model
%###########################################

\section{Model}\label{sec:model}

\subsection{Physical realisation}

\begin{figure}
    \centering
    \includegraphics{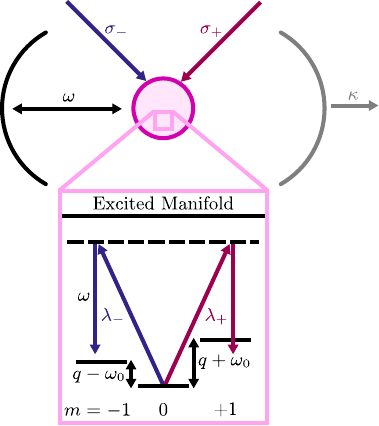}
    \caption{Illustration of the cavity-assisted Raman transition scheme that may be used to implement the spin-1 Dicke model. Atoms confined to a cavity with frequency $\omega$ are driven by two lasers with $\sigma_+$ and $\sigma_-$ polarizations. These combine with the cavity mode to drive Raman transitions, with coupling strengths $\lambda_+$ and $\lambda _-$, respectively, between the $m=0$ sublevel and the $m=\pm1$ sublevels, which have energies $q\pm\omega_0$. Cavity decay has a rate of $\kappa$.}
    \label{fig:schematic}
\end{figure}

In this work we consider an ensemble of $N$ identical spin-1 atoms coupled uniformly to an optical cavity mode. The Dicke model can be realised in such a system by driving cavity-assisted Raman transitions between 
atomic hyperfine ground-state sublevels \cite{dimer}. This is a well-established scheme, used previously in experiments to realise the traditional two-level Dicke model \cite{baum,klinder,Zhiqiang2018}, and adapted to spin-1 alkali atoms such as $^{87}$Rb \cite{zhiqiang17,masson2017}. 

The system schematic is illustrated in Fig. \ref{fig:schematic}. Two external lasers of $\sigma_+$ and $\sigma_-$ polarizations, respectively, are used to drive the atoms to an excited state manifold, while changing their angular momentum along the $z$-direction. They can subsequently emit a photon into the $\pi$ polarized cavity mode, returning to a ground state with a different $m$ quantum number. If the laser and cavity frequencies are detuned far from the atomic transition frequency, the excited states remain sparsely populated and transitions effectively occur directly between the ground states with respective coupling strengths of $\lambda_+$ and $\lambda_-$. By controlling the driving laser powers one can tune the Raman transition rates up and down the energy ladder and, thus, independently change the co-rotating and counter-rotating interaction strengths. In addition, this use of Raman transitions allows us to safely neglect spontaneous emission terms, such that cavity decay is the only relevant dissipative mechanism.

We also allow for a tunable energy level structure by taking advantage of Zeeman shifts, which split the otherwise degenerate magnetic sublevels. More specifically, we consider independently-controlled linear and quadratic Zeeman shifts, whose strengths we denote by $\omega_0$ and $q$ respectively. The $m=\pm1$ sublevels are shifted by $q\pm\omega_0$; hence, by allowing $q$ and $\omega_0$ to vary independently we achieve a tailored energy level structure. These shifts can be realized physically with static magnetic fields, or alternatively by inducing Stark shifts of the atomic levels through additional off-resonant laser fields \cite{you2017}.

\subsection{Hamiltonian}

Treatment of the traditional spin-1/2 Dicke model hinges on the mapping between a collection of \textit{individual} atomic spins and a single \textit{collective} atomic spin, i.e., from $N$ spin-1/2 atoms to a single `atom' with spin $N$/2. This is immensely helpful in reducing the effective size of the system's Hilbert space.

However, this mapping is not generally possible with the spin-1 Dicke model, whose Hamiltonian (with $\hbar=1$) is
\begin{widetext}
    \begin{equation}
        \h{H} = \omega\h{a}^\dag\h{a} + \sum_{n=1}^N\left[ \omega_0\h{S}_{n,z} + q\h{S}^2_{n,z} + \frac{\lambda_-}{\sqrt{2N}}(\h{a}\h{S}_{n,+}+\h{a}^\dag\h{S}_{n,-}) + \frac{\lambda_+}{\sqrt{2N}}(\h{a}\h{S}_{n,-}+\h{a}^\dag\h{S}_{n,+}) \right]\,,
    \end{equation}
\end{widetext}
\noindent 
where $\omega$ and $\omega_0$ are the (effective) cavity and atomic frequencies, respectively, $\lambda_\pm$ are the atom-cavity-mode coupling strengths, 
and $\h{S}_{n,(\cdot)}$ denotes the $n$th individual atomic spin operator. With the collective-spin (or angular momentum) operators defined as
\begin{equation}
    \h{S}_{(\cdot)}\equiv\sum_{n=1}^N\h{S}_{n,(\cdot)}\,,
\end{equation}
\noindent it becomes evident that they cannot represent the Hamiltonian term $\sum_{n=1}^N \h{S}^2_{n,z}$ because it includes the \textit{squares} of individual spin operators. Hence, unless $q=0$, this mapping cannot be employed.

Instead, we make use of the Jordan-Schwinger map \cite{schwinger} to express the atomic degrees of freedom in terms of three harmonic oscillator modes (one for each atomic energy level). If the respective annihilation operators of these modes are $\h{b}_+$, $\h{b}_-$, and $\h{b}_0$, we have
\begin{align}
    \sum_{n=1}^N \h{S}^2_{n,z} &= \sum_{i,j=0,+,-}\h{b}_i^\dag\h{b}_j\bra{i}_n\h{S}^2_{n,z}\ket{j}_n\nonumber\\
    &=\h{b}^\dag_+\h{b}_+ + \h{b}^\dag_-\h{b}_-\,,
\end{align}
\noindent where $\ket{i}_n$ refers to the individual magnetic sublevels $\ket{m=0,\pm1}$ of the $n$th atom. Similarly, we have
\begin{align}
    \h{S}_z=\sum_{n=1}^N \h{S}_{n,z} &= \h{b}^\dag_+\h{b}_+ - \h{b}^\dag_-\h{b}_-\,,\\
    \h{S}_-=\sum_{n=1}^N \h{S}_{n,-} &= \sqrt{2}(\h{b}_0^\dag\h{b}_+ + \h{b}_-^\dag\h{b}_0)\,.
\end{align}

Note that by using the regular annihilation operator algebra, namely
\begin{align}
    [\h{b}_i,\h{b}_j]&=0\,, & [\h{b}_i,\h{b}_j^\dag]=\delta_{ij}\,,
\end{align}
\noindent one can recover the collective spin operator algebra,
\begin{align}
    [\h{S}_z,\h{S}_\pm]&=\pm\h{S}_\pm\,, & [\h{S}_+,\h{S}_-]&=2\h{S}_z\,.
\end{align}

The Hamiltonian can then be rewritten as
\begin{widetext}
    \begin{equation}\label{eq:hamiltonian}
        \h{H}=\omega\h{a}^\dag\h{a} + (q+\omega_0)\h{b}_+^\dag\h{b}_+ + (q-\omega_0)\h{b}_-^\dag\h{b}_-+\frac{1}{\sqrt{N}}(\lambda_+\h{a}+\lambda_-\h{a}^\dag)(\h{b}_0^\dag\h{b}_+ + \h{b}_-^\dag\h{b}_0)+\frac{1}{\sqrt{N}}(\lambda_-\h{a}+\lambda_+\h{a}^\dag)(\h{b}_0^\dag\h{b}_- + \h{b}_+^\dag\h{b}_0)\,.
    \end{equation}
\end{widetext}

Lastly, we account for dissipation through cavity decay, meaning that the evolution of the system density operator is given by the master equation
\begin{equation}\label{eq:master_equation}
    \frac{d\h{\rho}}{dt}=-i[\h{\rho},\h{H}]+\kappa(2\h{a}\h{\rho}\h{a}^\dag-\h{a}^\dag\h{a}\h{\rho}-\h{\rho}\h{a}^\dag\h{a})\,,
\end{equation}
\noindent where $\h{\rho}$ is the system's density operator and $\kappa$ is the cavity field decay rate.

\subsection{Semiclassical equations of motion}

In this work we are interested in taking the semiclassical limit of $N\to\infty$ of the model given by Eqs. (\ref{eq:hamiltonian}) and (\ref{eq:master_equation}). In this limit, quantum fluctuations can be ignored and non-commuting operators can be replaced by commuting complex numbers. This allows operator expectations to be factorised, e.g., $\expv{\h{a}^\dag\h{a}}=\expv{\h{a}^\dag}\expv{\h{a}}$, and, hence, the equations of motion for the expectations become nonlinear.

Introducing the complex variables
\begin{align}
    a&\equiv\frac{\expv{\h{a}}}{\sqrt{N}}\,, & b_\pm&\equiv\sqrt{\frac{2}{N}}\expv{\h{b}_\pm}\,, & b_0&\equiv\frac{\expv{\h{b}_0}}{\sqrt{N}}\,,
\end{align}
\noindent we obtain the semiclassical equations of motion
\begin{align}
    \dot{a}&=-(\kappa+i\omega)a-i\lambda_-(b_0^*b_++b_-^*b_0)\nonumber\\
    &\hspace{2.2cm}-i\lambda_+(b_+^*b_0+b_0^*b_-)\,,\label{eq:eoms1}\\
    \dot{b}_+&=-i(q+\omega_0)b_+-i(\lambda_-a+\lambda_+a^*)b_0\,,\label{eq:dbp}\\
    \dot{b}_-&=-i(q-\omega_0)b_--i(\lambda_+a+\lambda_-a^*)b_0\,,\label{eq:dbm}\\
    \dot{b}_0&=-i(\lambda_-a+\lambda_+a^*)b_--i(\lambda_+a+\lambda_-a^*)b_+\,.\label{eq:eoms2}
\end{align}
\noindent They are the central object of study in this work. Namely, our interest lies in solutions of this system of ordinary differential equations and their bifurcations, which correspond to (quantum) phase transitions.

Equations (\ref{eq:eoms1})--(\ref{eq:eoms2}) have a number of symmetries we can exploit. Specifically, $\mathbb{Z}_2$, U(1), and parameter exchange symmetries, which are expressed, respectively, as invariance of Eqs. (\ref{eq:eoms1})--(\ref{eq:eoms2}) under the transformations
\begin{align}
    T_{\mathbb{Z}_2}: (a,b_+,b_-)&\to(-a,-b_+,-b_-)\,,\\
    T_{U(1)}: (b_+,b_-,b_0)&\to(e^{i\theta}b_+,e^{i\theta}b_-,e^{i\theta}b_0)\,,\label{eq:phase_inv}\\
    T: (b_+,b_-,\lambda_+,\lambda_-,\omega_0)&\to(b_-,b_+,\lambda_-,\lambda_+,-\omega_0)\,,
\end{align}
\noindent where $\theta$ is a constant phase shift. In addition to these symmetries, we must have
\begin{equation}\label{eq:num_con}
    |b_+|^2+|b_-|^2+|b_0|^2=1
\end{equation}
\noindent as a consequence of atom number conservation.

Given $a$, $b_+$, $b_-$, and $b_0$ are all complex, the phase space of Eqs. (\ref{eq:eoms1})--(\ref{eq:eoms2}) is $\mathbb{C}^4\equiv\mathbb{R}^8$. However, Eq. (\ref{eq:num_con}) constrains the atomic dynamics to the sphere $\mathbb{S}^5$. Furthermore, phase invariance (under the transformation $T_{U(1)}$) allows one to reduce the atomic dynamics to the lower-dimensional sphere $\mathbb{S}^4$ by applying an appropriate phase rotation. (A lower-dimensional analogy is moving to a rotating frame for dynamics on the two-dimensional sphere $\mathbb{S}^2$ in order to reduce them to the circle $\mathbb{S}$.) Therefore, altogether, our system's effective phase space is $\mathbb{R}^2\times\mathbb{S}^4$.

\subsubsection*{The normal phase}

Understanding the emergence of superradiance in Dicke models requires an understanding of their dark states. These are steady-states of the system with an empty cavity, which together comprise the normal phase. In the traditional two-level Dicke model there are two states belonging to the normal phase, with either a completely excited or completely de-excited atomic ensemble \cite{garraway2011,stitely}. However, in our spin-1 model the normal phase is nontrivial and its dark states can be separated into two categories.

The first are states where all atoms are in the $m=0$ sublevel. Semiclassically, with reference to Eqs. (\ref{eq:eoms1})--(\ref{eq:eoms2}), these states correspond to the equilibria $(a,b_+,b_-,b_0)=(0,0,0,e^{i\eta})$, with an arbitrary phase $\eta$. As we will see in the following sections, these equilibria are responsible for the emergence of equilibrium superradiance in the V configuration and, therefore, for other dynamics that follow equilibrium superradiance. Given the equilibria's association with the $m=0$ sublevel, we refer to them as N$_0$.

Conversely, the second group of dark states are the so-called `inverted states' \cite{lin2022} where no atoms are in the $m=0$ sublevel. That is, a portion is in the $m=+1$ sublevel and the remainder is in the $m=-1$ sublevel. Semiclassically, they are a family of trajectories which lie on the surfaces of 4-tori. They are paramaterised by $r_+=|b_+|$ and $r_-=|b_-|$, and given by $(a,b_+,b_-,b_0)=(0,r_+e^{i(q+\omega_0)t},r_-e^{i(q-\omega_0)t},0)$. Note that, because of Eq. (\ref{eq:num_con}) (number conservation), the members of this family with $r_+=1$ or $r_-=1$ are the periodic orbits $(a,b_+,b_-,b_0)=(0,e^{i(q+\omega_0)t},0,0)$ and $(0,0,e^{i(q-\omega_0)t},0)$, respectively. Physically, they represent states where all the atoms are in either the $m=+1$ or $m=-1$ sublevel, and, hence, we refer to them as N$_+$ and N$_-$.

\subsubsection*{Stereographic projection and phase rotation}\label{sec:proj}

Throughout this work we make extensive use of the numerical continuation package AUTO-07P \cite{doedelauto07p}. It uses pseudo-arclength continuation to follow equilibria and periodic orbits as parameters are varied, track their stability, and then follow the bifurcations they undergo in multiple parameters. Because AUTO assumes that the system phase space is $\mathbb{R}^n$, and our phase space is $\mathbb{R}^2\times\mathbb{S}^4$, we need to recast Eqs. (\ref{eq:eoms1})--(\ref{eq:eoms2}) with the restriction given by Eqs. (\ref{eq:phase_inv}) and (\ref{eq:num_con}). We do so by means of a stereographic projection in combination with a phase rotation. Defining $r_0=|b_0|$ and $\phi_0=\arg(b_0$), this amounts to the transformation
\begin{equation}
    (b_+,b_-,b_0)\to(x,y)=\left(\frac{b_+e^{-i\phi_0}}{1-r_0},\frac{b_-e^{-i\phi_0}}{1-r_0}\right)\,,
\end{equation}
\noindent where $x$ and $y$ are complex.

The resulting equations of motion for $a$, $x$, and $y$ are
\begin{widetext}
    \begin{align}
        \dot{a}&=-(\kappa+i\omega)a-\frac{2i(|x|^2+|y|^2-1)}{(|x|^2+|y|^2+1)^2}\left[\lambda_-(x+y^*)+\lambda_+(x^*+y)\right]\,,\label{eq:sys3}\\
        \dot{x}&=-i(q+\omega_0)x-\frac{iA}{2}(|x|^2+|y|^2-1)+\frac{ix}{|x|^2+|y|^2-1}\left[A(x^*+y)+A^*(x+y^*)\right]+\frac{ix}{2}\left[A(x^*-y)+A^*(x-y^*)\right]\,,\\
        \dot{y}&=-i(q-\omega_0)y-\frac{iA^*}{2}(|x|^2+|y|^2-1)+\frac{iy}{|x|^2+|y|^2-1}\left[A(x^*+y)+A^*(x+y^*)\right]-\frac{iy}{2}\left[A(x^*-y)+A^*(x-y^*)\right]\,, \label{eq:sys4}
    \end{align}    
\end{widetext}
\noindent where we write $A\equiv\lambda_-a+\lambda_+a^*$ for convenience. Note that the system given by Eqs. (\ref{eq:sys3})--(\ref{eq:sys4}) in $\mathbb{C}^3\equiv\mathbb{R}^6$ has no constraints and can, hence, readily be investigated with AUTO. Similarly to the un-transformed system in Eqs. (\ref{eq:eoms1})--(\ref{eq:eoms2}), it possesses both $\mathbb{Z}_2$ and parameter exchange symmetries, under the transformations
\begin{align}
    \mathcal{T}_{\mathbb{Z}_2}:(a,x,y)&\to(-a,-x,-y)\,,\label{eq:z2_sym}\\
    \mathcal{T}:(x,y,\lambda_+,\lambda_-,\omega_0)&\to(y,x,\lambda_-,\lambda_+,-\omega_0)\,.\label{eq:xy_par_ex}
\end{align}
\noindent In addition, N$_0$ is represented by the equilibrium $(a,x,y)=(0,0,0)$, which is particularly convenient because N$_0$ is responsible for the transition to superradiance in the V configuration.

However, the coordinate transformation comes with two caveats: the first is that all points with $r_0=1$ are projected infinitely far away in the $(x,y)$ hyperplane, and the second is that all points with $r_0=0$ have an undefined phase $\phi_0$. The latter set of points, including the normal phase tori as well as N$_+$ and N$_-$, are transformed to the unit $(x,y)$-circle $|x|^2+|y|^2=1$, where Eqs. (\ref{eq:sys3})--(\ref{eq:sys4}) are not defined. Hence, in order to describe their dynamics in a suitably reduced phase space (i.e., not in $\mathbb{C}^4\equiv\mathbb{R}^8$) the phase rotation and stereographic projection can be done with respect to $r_\pm=|b_\pm|$ and $\phi_\pm=\arg(b_\pm)$ instead of $r_0$ and $\phi_0$. In Appendix \ref{apdx:proj} we do so with respect to $r_-$ and $\phi_-$, and give the resulting equations of motion. These have analogous issues when $r_-=1$ and $r_-=0$, but they suffice in describing the dynamics of N$_-$, which is mapped to the origin. Hence, in order to give a full dynamical picture involving both N$_+$ and N$_-$, we make use of the different coordinate transformations.

%###########################################
%                Symmetric V
%###########################################

\section{Symmetric V Configuration}\label{sec:sym_v}

The symmetric V configuration corresponds to setting the linear Zeeman shift to zero, and keeping the quadratic shift positive, i.e., setting $\omega_0=0$ and $q>0$. We wish to study the effect of unbalanced co-rotating and counter-rotating coupling terms, while fixing the structure of the atomic levels. Experimentally, this corresponds to changing the powers and frequencies of the driving lasers, while fixing the static magnetic fields (or equivalent mechanism) that determine the level structure. Therefore, we fix $\kappa$, $\omega$, and $q$, while varying $\lambda_+$ and $\lambda_-$. More specifically, we set $\kappa=\omega=q=1$ throughout this section.

\begin{figure}
    \centering
    \includegraphics{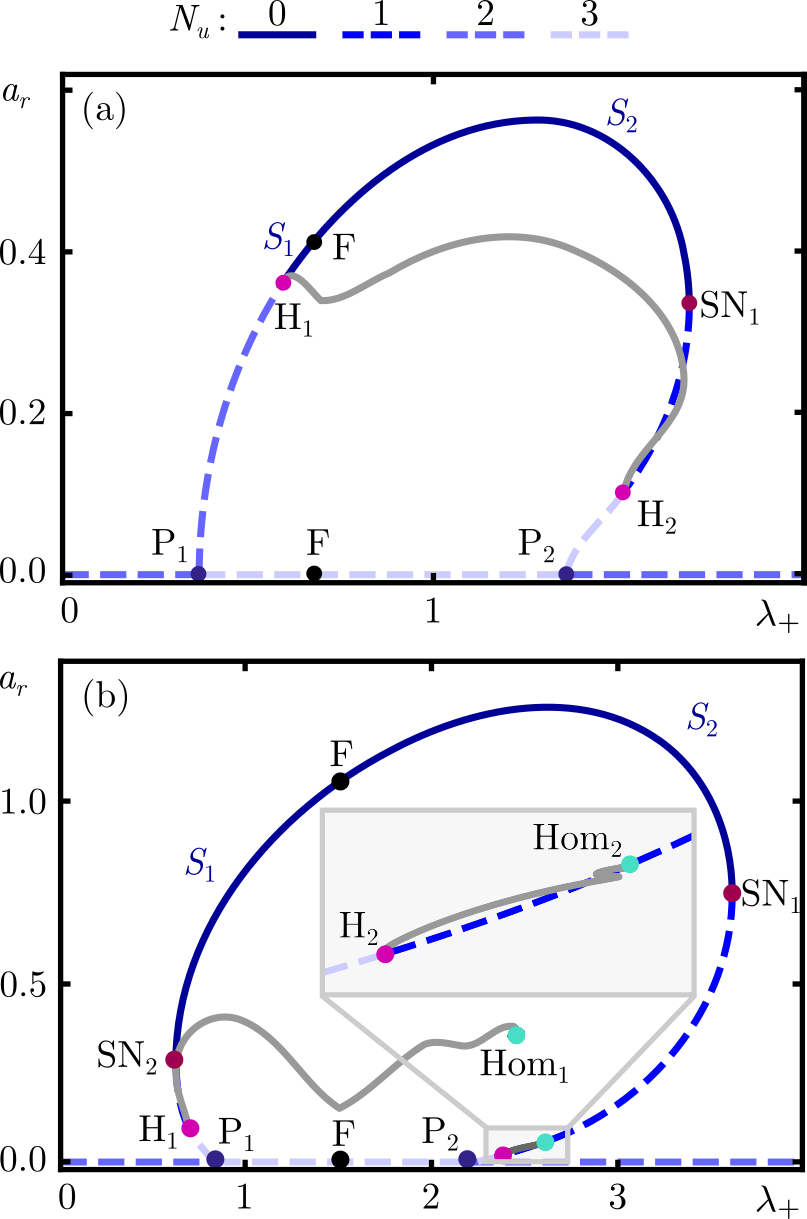}
    \caption{One-parameter bifurcation diagrams in $\lambda_+$ for (a) $\lambda_-=0.7$, and (b) $\lambda_-=1.5$, shown in terms of $a_r=\Re(a)$. Solid and dashed blue curves denote stable and unstable equilibria, respectively. The number of unstable directions $N_u$ is indicated according to the key above. Pitchfork, saddle-node, Hopf and superradiant-flip bifurcations are labelled P, SN, H, and F, respectively. Note that we only show the positive $a_r$ equilibrium branch, since the negative one is simply related by the $\mathbb{Z}_2$ symmetry. Unstable periodic orbit branches are shown in gray as represented by the maximum value of $a_r$ along each orbit. Their homoclinic bifurcations are labelled Hom$_1$ and Hom$_2$.}
    \label{fig:sym_v_one_par}
\end{figure}

We begin by fixing $\lambda_-$ and following N$_0$ and its stability as $\lambda_+$ is varied. Figure ~\ref*{fig:sym_v_one_par} contains two resultant bifurcation diagrams, each showing branches of equilibria as functions of $\lambda_+$ together with their stabilities. Any system variable may be chosen to represent the locations of the equilibria; we choose $a_r=\text{Re}(a)$ here.

In both panels of Fig. \ref{fig:sym_v_one_par} we see symmetry-breaking pitchfork bifurcations P$_1$ and P$_2$ of N$_0$, which give rise to two superradiant equilibria each. These bifurcations occur whenever the Jacobian of the system, evaluated at N$_0$, has a zero eigenvalue, or alternatively when its determinant is zero, i.e., when
\begin{align}\label{eq:xy_origin_pf}
    0&=(\kappa^2+\omega^2)(\omega_0^2-q^2)^2+4q^2(\lambda_+^2-\lambda_-^2)^2\nonumber\\
    &\hspace{3cm}+4\omega q(\lambda_+^2+\lambda_-^2)(\omega_0^2-q^2)\,.
\end{align}
\noindent These bifurcations may be super- or subcritical; in Fig.~\ref*{fig:sym_v_one_par}(a), the initial pitchfork P$_1$ is supercritical, while in Fig. \ref*{fig:sym_v_one_par}(b) it is subcritical.

The superradiant equilibria that emerge as a result of P$_1$ and P$_2$ are always found on the hyperplane given by $x=y^*$; see Appendix \ref{apdx:eqs} for details. Therefore, one equilibrium has a positive $x_i=\Im(x)$ and a negative $y_i=\Im(y)$, while the other has a positive $y_i$ and a negative $x_i$, by virtue of the $\mathbb{Z}_2$ symmetry given by Eq. (\ref{eq:z2_sym}). Physically, lying on the $x=y^*$ hyperplane implies that the superradiant equilibria correspond to states where the $m=+1$ and $m=-1$ sublevels are equally populated. 

These equilibria undergo additional bifurcations, including the Hopf bifurcations H$_1$ and H$_2$, the saddle-node bifurcations SN$_1$ and SN$_2$, as well as a transition we dub the \textit{superradiant-flip}; it occurs when $\lambda_+=\lambda_-$ and is marked F in Fig. \ref{fig:sym_v_one_par}. Together, the bifurcations delineate the regions of superradiant stability labelled $S_1$ and $S_2$ in Figs. \ref{fig:sym_v_one_par}(a) and (b). For example, in Fig. \ref{fig:sym_v_one_par}(a), $S_1$ is bounded by H$_1$ and F, while $S_2$ is bounded by F and SN$_1$.

\subsubsection*{Dynamics at the superradiant-flip transition}

A unique scenario arises at the superradiant-flip point as a result of additional symmetry. Since $\omega_0=0$ in the symmetric V configuration, the parameter exchange symmetry given by Eq. (\ref{eq:xy_par_ex}) is expressed as invariance under the transformation $\mathcal{T}_V(x,y,\lambda_+,\lambda_-)=\mathcal{T}(x,y,\lambda_+,\lambda_-,0)$, where
\begin{equation}
    \mathcal{T}_V:(x,y,\lambda_+,\lambda_-)\to(y,x,\lambda_-,\lambda_+)\,.
\end{equation}
Therefore, at the superradiant-flip point, when $\lambda_+=\lambda_-\equiv\lambda$, this reduces to $\mathbb{Z}_2$ symmetry which is expressed as invariance under the transformation $\hat{\mathcal{T}}_V(x,y)=\mathcal{T}_V(x,y,\lambda,\lambda)$, where
\begin{equation}
    \hat{\mathcal{T}}_V:(x,y)\to(y,x)\,,\label{eq:z2}
\end{equation}
\noindent whose symmetry subspace are all the points lying on the $x=y$ hyperplane. This includes the superradiant equilibria, as demonstrated in Appendix \ref{apdx:eqs}, meaning these lie at the intersection of the $x=y$ and the $x=y^*$ hyperplanes. In other words, their $x$ and $y$ coordinates are both real and equal. This occurs because their locations on the $x=y^*$ hyperplane are exchanged in the course of the superradiant-flip transition: with reference to the previous section, the equilibrium with positive $x_i$ and a negative $y_i$ becomes the equilibrium with positive $y_i$ and a negative $x_i$, and vice versa. At the point of exchange, i.e., at the superradiant-flip, the $x_i$ and $y_i$ of both equilibria are equal to zero.

\begin{figure}
    \centering
    \includegraphics{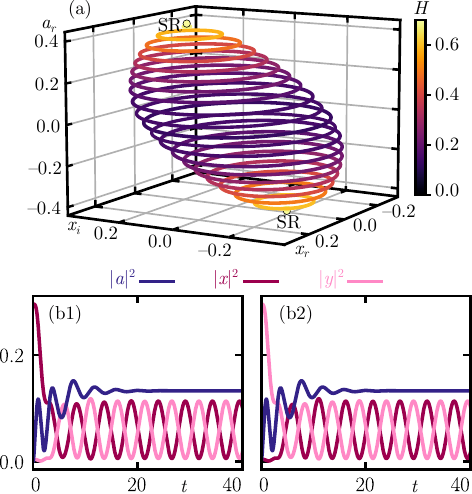}
    \caption{Panel (a) shows the periodic orbit foliation in $(x_r,x_i,a_r)$-space at the superradiant-flip bifurcation, for $\lambda=0.7$. The energy $H$ of each periodic orbit is represented by its colour. The superradiant equilibria are labelled SR and lie at the top and bottom ends of the resultant surface. Panels (b1) and (b2) show $|a|^2$, $|x|^2$, and $|y|^2$ (as per the legend) for trajectories with initial conditions $(a,x,y)=(0,0.544,0.065)$ and $(a,x,y)=(0,0.065,0.544)$, respectively.}
    \label{fig:flip}
\end{figure}

In addition, both superradiant equilibria and N$_0$ acquire a pair of conjugate imaginary eigenvalues. However, as seen in Fig. \ref{fig:sym_v_one_par}, the stability of those equilibria is not affected. Instead, at the transition infinitely-many periodic orbits exist, which foliate a region of phase space that connects both superradiant equilibria. Figure \ref{fig:flip}(a) shows a selection of periodic orbits on the surface for $\lambda=0.7$, in projection onto $(x_r,x_i,a_r)$-space. The energy per atom, $H\equiv\expv{\h{H}}/N$, along each orbit is conserved and given by
\begin{align}
    H&=\omega|a|^2+\frac{q}{2}(|b_+|^2+|b_-|^2)\nonumber\\
    &\hspace{0.8cm}+\sqrt{2}\lambda a_r[(b_+^*b_0+b_0^*b_-)+(b_0^*b_++b_-^*b_0)]\nonumber\\
    &=\omega|a|^2+\frac{q(|x|^2+|y|^2)}{(1+|x|^2+|y|^2)^2}\nonumber\\
    &\hspace{0.8cm}+2\sqrt{2}\lambda a_r(x_r+y_r)\frac{1-|x|^2-|y|^2}{(1+|x|^2+|y|^2)^2}\,,\label{eq:H}
\end{align}
\noindent where $y_r\equiv\Re(y)$. As shown in Fig. \ref{fig:flip}(a), where this energy is represented by colour, $H$ increases with $|a_r|$ until it attains its maximum at the superradiant equilibria.

Because of the symmetry given by Eq. (\ref{eq:z2}), every orbit has a symmetric counterpart given by application of $\hat{\mathcal{T}}_V$. In other words, the surface formed by the periodic orbits in Fig. \ref{fig:flip}(a) will look identical if projected with respect to $y_r$ and $y_i$ instead of $x_r$ and $x_i$. Figures \ref{fig:flip}(b1) and (b2) show two trajectories that converge to symmetric-counterpart periodic orbits on the superradiant surface. Note that the cavity population $|a|^2$ tends to the same constant value in both trajectories, as is typical for equilibrium superradiance. However, the atomic variables $x$ and $y$ oscillate sinusoidally and perfectly out of phase. Hence, the pair of symmetric counterpart periodic orbits, with $x$ and $y$ exchanged as seen in Figs. \ref{fig:flip}(b1) and (b2), must be out of phase with each other. Since Eq. (\ref{eq:H}) is invariant under the transformations $\mathcal{T}_{\mathbb{Z}_2}$ and $\hat{\mathcal{T}}_V$, the energy $H$ does not uniquely define each orbit on the surface: for every periodic orbit there exist three others that have the same energy, given by exchange of $x$ and $y$ and/or inversion of their signs. However, the energy \textit{does} parameterize the surface since it uniquely defines the set of four symmetry-related periodic orbits. Therefore, and because each periodic orbit forms a one-dimensional closed curve in phase space, the superradiant surface as a whole is, in fact, two-dimensional.

The presence of such energy-conserving trajectories is non-trivial, given they appear when dissipation is present. The physical explanation for their existence lies in a fine balance of energy exchange between the atoms, the cavity, and the environment. Since $\lambda_+=\lambda_-$ the atoms are being driven equally strongly up and down the energy ladder. In addition, since the level structure is symmetric, transition rates to and from the $m=0$ sublevel are equal. Hence, the population of the $m=0$ sublevel is constant, and photons are emitted into the cavity at a constant rate. Conversely, due to cavity decay, photons are lost to the environment at a constant rate. Thus, if a balance is struck between these processes, the total energy of the system can remain constant despite its oscillatory behaviour.

A similar transition, the \textit{pole-flip}, was observed in the two-level Dicke model \cite{stitely}. It is also facilitated by an atom-cavity-environment energy balance and results in a foliation of phase space by energy-conserving periodic orbits. That being said, there are differences in the dimensionality and symmetries between these two systems, which result in topologically different superradiant surfaces. Nevertheless, the similarity between the pole-flip and superradiant-flip suggests that this type of transition is a general feature of driven-dissipative atomic systems.

\subsubsection*{Bifurcations of periodic orbits}

Other periodic orbits appear when $\lambda_+\neq\lambda_-$ through a pair of Hopf bifurcations H$_1$ and H$_2$ that occur along the superradiant branches in both panels of Fig. \ref{fig:sym_v_one_par}. In Fig.~\ref{fig:sym_v_one_par}(a) a single periodic orbit branch originates at H$_1$ and terminates at H$_2$, while in Fig.~\ref{fig:sym_v_one_par}(b) the two Hopf bifurcations give rise to two separate branches, which end at a pair of homoclinic bifurcations, labelled Hom$_1$ and Hom$_2$, respectively.

\begin{figure}
    \centering
    \includegraphics[width=\columnwidth]{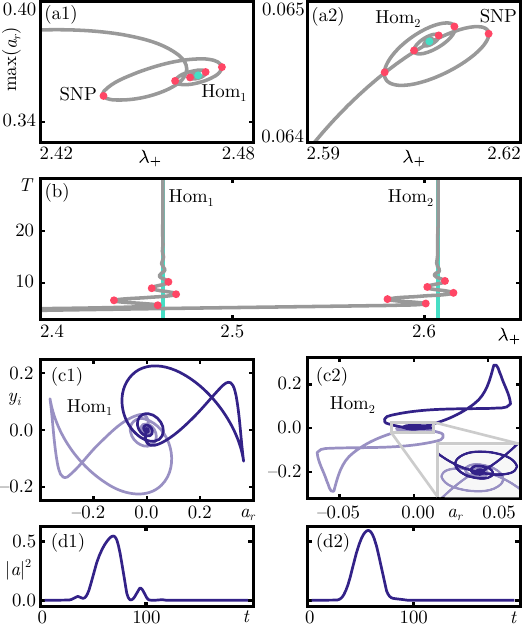}
    \caption{Homoclinic bifurcations for $\lambda_-=1.5$. Panels (a1) and (a2) are enlargements of the bifurcation diagram near Hom$_1$ and Hom$_2$, respectively. Several saddle-node bifurcations of periodic orbits (SNP) are marked in red, and the homoclinic points are marked in cyan. Panel (b) shows the period $T$ as a function of $\lambda_+$ of the periodic orbits as they approach the homoclinic points, where SNP points are marked as in panels (a1) and (a2). Panels (c1) and (c2) show the homoclinic orbits themselves, in projection onto the $(a_r,y_i)$-plane, where the two asymmetric counterparts are distinguished by colour. Panels (d1) and (d2) show the cavity population $|a|^2$ as a function of time along each homoclinic trajectory.}
    \label{fig:sym_hom}
\end{figure}

Fig.~\ref*{fig:sym_hom} further illustrates the bifurcation diagram from Fig. \ref*{fig:sym_v_one_par}(b) in the vicinity of the homoclinic bifurcations. In panels (a1) and (a2) we see the bifurcation diagram near the points Hom$_1$ and Hom$_2$, respectively; in both cases the branch of periodic orbits spirals toward the homoclinic bifurcation. This occurs on a much smaller parameter scale than that of Fig. \ref*{fig:sym_v_one_par}(b). 

A saddle-node bifurcation of periodic orbits, labelled SNP, occurs at every turning point, where two periodic orbits appear and move apart, or collide and disappear. The SNP points accumulate on the homoclinic bifurcation points Hom$_1$ and Hom$_2$ while the period of each orbit grows beyond bound; this is illustrated by Fig.~\ref*{fig:sym_hom}(b), where the SNP points are also marked.

At the points Hom$_1$ and Hom$_2$, when the period becomes `infinite', one finds a pair of asymmetric homoclinic orbits exist to the origin. These are shown in Figs.~\ref{fig:sym_hom}(c1) and (c2), in projection onto the $(a_r,y_i)$-plane, where they can be seen leaving the origin and spiralling back to it. These homoclinic orbits are notable because they are trajectories that originate from and return to the same point in phase space, namely N$_0$. Hence, along such a trajectory, one would observe an initially empty cavity at N$_0$, followed by a burst of photons as the trajectory leaves the $a=0$ plane, with an eventual decay of the cavity population as N$_0$ is approached again. This is illustrated in Figs. \ref{fig:sym_hom}(d1) and (d2), which show the cavity population as a time series along the homoclinic orbits Hom$_1$ and Hom$_2$.

\subsubsection*{Stability of the normal phase tori, and phase diagram}

\begin{figure}
    \centering
    \includegraphics{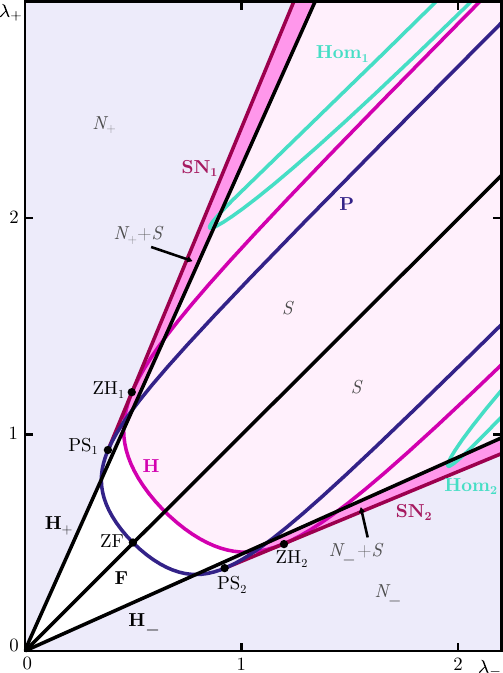}
    \caption{Phase diagram for the symmetric V configuration, with $\kappa=\omega=q=1$ and $\omega_0=0$. In the regions labelled $N_\pm$ (blue shading) only N$_\pm$ are stable. Similarly, in the region $S$ (pink shading) only superradiant equilibria are stable. In the regions labelled $N_\pm+S$ there is normal-superradiant bistability. Codimension-one pitchfork (\textbf{P}), saddle-node (\textbf{SN}), Hopf (\textbf{H}), homoclinic (\textbf{Hom}), and superradiant-flip (\textbf{F}) bifurcations are shown as curves and labelled accordingly. Note that the curves \textbf{H}$_{\pm}$ represent Hopf bifurcations of N$_\pm$. Codimension-two pitchfork-saddle-node (PS), zero-Hopf (ZH), and zero-flip (ZF) bifurcations are plotted as points.}
    \label{fig:sym_diag}
\end{figure}

The last objects left to consider are the normal phase tori N$_+$ and N$_-$. One can show that they only change stability through Hopf bifurcations, when
\begin{align}
    0&=\pm[\kappa^2+\omega^2+(q\pm\omega_0)^2](\lambda_-^2-\lambda_+^2)\nonumber\\&\hspace{3.5cm}+2\omega(q\pm\omega_0)(\lambda_-^2+\lambda_+^2)\,,
\end{align}
\noindent where the choice of $\pm$ above corresponds, respectively, to N$_+$ or N$_-$. This allows us to construct the phase diagram in Fig.~\ref*{fig:sym_diag}, which shows the location of each bifurcation as a curve in the $(\lambda_-,\lambda_+)$-plane. These curves delineate regions where the relevant objects are stable and hence determine where the normal or superradiant phases are experimentally observable. For example, N$_\pm$ are stable in the regions $N_\pm$, and superradiant equilibria are stable in the region $S$. Regions of normal-superradiant bistability, where N$_+$ or N$_-$ are stable alongside a superradiant equilibrium, are also present and labelled $N_\pm+S$.

The bifurcation curves intersect several times in so-called \textit{codimension-two} bifurcations. These often involve changes in criticality along the bifurcation curves, the emergence of additional bifurcation curves, a combination of the two, or the implication of complicated dynamics nearby \cite{kuznetsov}. Note that not every intersection is a codimension-two bifurcation, since they do not necessarily involve the same dynamical object. In our case, codimension-two pitchfork-saddle-node bifurcations labelled PS$_1$ and PS$_2$ occur at the tangential intersection of the pitchfork and saddle-node curves \textbf{P}, \textbf{SN}$_\textbf{1}$, and \textbf{SN}$_\textbf{2}$. These mark the change in criticality of the pitchfork bifurcations P$_1$ and P$_2$ from Fig. \ref{fig:sym_v_one_par}. The Hopf and saddle-node curves also intersect tangentially, at two zero-Hopf points labelled ZH$_1$ and ZH$_2$, where the superradiant equilibria simultaneously have a zero eigenvalue and a pair of imaginary eigenvalues. Similarly to PSN$_1$ and PSN$_2$, the points ZH$_1$ and ZH$_2$ are associated with a change of criticality: in this case of the Hopf bifurcations along \textbf{H}. Hence, ZH$_1$ and ZH$_2$ mark a change in the delineation of the region $S$. Lastly, the pitchfork and superradiant-flip bifurcation curves also intersect, but transversally, at the zero-flip point ZF. At this intersection, the normal phase equilibrium point has a zero-eigenvalue and a pair of imaginary eigenvalues, much like a zero-Hopf bifurcation. However, unlike that bifurcation, it does not mark a change in the regions of stability or a change in criticality.

%###########################################
%                Asymmetric V
%###########################################

\section{Asymmetric V Configuration}\label{sec:asym_v}

As a consequence of introducing an effective quadratic Zeeman shift to our model, we enable tunability of the atomic energy level configuration. We are, therefore, not restricted to studying the symmetric V configuration. As we will see, level asymmetry provides a mechanism for complicated dynamics to emerge. In particular, the previous energy balance at the superradiant-flip bifurcation breaks and stable oscillations develop, precisely because the energy levels are no longer degenerate and imbalanced transitions occur between them.

In terms of our system parameters, introducing a level asymmetry to the V configuration amounts to setting $\omega_0$ to a nonzero value and maintaining $q>|\omega_0|$. In order to keep the energy level splitting on the same order of magnitude as the cavity frequency $\omega$, we impose the additional constraint that $q+\omega_0=1$. Therefore, we change the level asymmetry by varying $q$ and $\omega_0$ simultaneously such that $q-\omega_0$ is changed, and $q+\omega_0$ remains fixed. 

\subsection{Bifurcations of equilibria and superradiant-flip breakdown}\label{sec:asym_v_eq}

\begin{figure}
    \centering
    \includegraphics[width=\columnwidth]{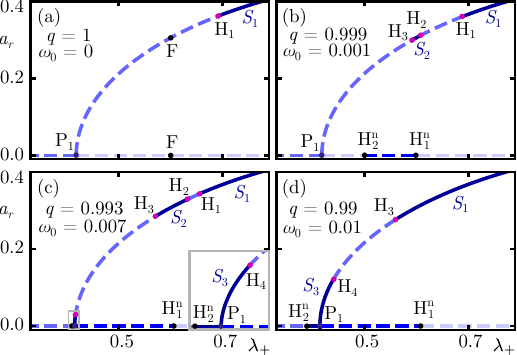}
    \caption{Bifurcation diagrams for $\lambda_-=0.6$ and increasing level asymmetry; the respective $q$ and $\omega_0$ values are given in each panel. Also shown are the superradiant-flip bifurcations F, pitchfork bifurcations P, superradiant Hopf bifurcations H, and normal-phase Hopf bifurcations H$^n$. Segments of stable superradiant equilibria are labelled S. The inset in panel (c) shows an enlargement near the pitchfork point P.}
    \label{fig:asym_lm=0.6}
\end{figure}

We now consider how the balance held at the superradiant-flip bifurcation breaks, once the degeneracy of the $m=-1$ and $m=+1$ sublevels is lifted, and how this process is expressed in bifurcations of equilibria. Figure ~\ref*{fig:asym_lm=0.6} shows one-parameter bifurcation diagrams for $\lambda_-=0.6$ near the superradiant-flip and pitchfork bifurcations, for a progressively more asymmetric level structure.

Fig. \ref{fig:asym_lm=0.6}(a), for $q=1$ and $\omega_0=0$, is a bifurcation diagram of the symmetric V configuration; compare with Fig.~\ref*{fig:sym_v_one_par}. The normal phase equilibrium point undergoes a supercritical pitchfork bifurcation P$_1$, thereby giving rise to two additional superradiant equilibria related by the system's $\mathbb{Z}_2$ symmetry. At $\lambda_+=\lambda_-=0.6$, these equilibria undergo the superradiant-flip bifurcation F, which leaves their stability unchanged. They then become stable at a Hopf bifurcation labelled H$_1$, and remain stable along the segment of the branch labelled $S_1$.

Introducing even a small level asymmetry, as in panel (b) for $q=0.999$ and $\omega_0=0.001$, changes this picture appreciably. Although P$_1$ and H$_1$ remain largely unaffected, the superradiant-flip bifurcation splits into the pair of superradiant Hopf bifurcations H$_2$ and H$_3$, as well as the pair of normal-phase Hopf bifurcations H$^n_1$ and H$^n_2$. As a result, a small window of stable superradiance, labelled $S_2$, opens along the branch segment between H$_2$ and H$_3$.

This window grows as the asymmetry is increased; in panel (c) $S_2$ is larger, since H$_2$ moves up the branch towards H$_1$, and H$_3$ moves down the branch towards P$_1$. We also see a new region of superradiant stability develop as a consequence of crossing a codimension-two pitchfork-Hopf bifurcation. This bifurcation occurs when P$_1$ and H$^n_2$ collide, and the normal phase equilibrium N$_0$ has both a zero eigenvalue and a pair of imaginary eigenvalues. As a result, a new superradiant Hopf bifurcation H$_4$ appears, as shown in panel (c). Moreover, a small segment of normal-phase stability emerges between H$^n_2$ and P$_1$, and, since P$_1$ is supercritical, this leads to another segment of superradiant stability, $S_3$.

Both $S_2$ and $S_3$ continue to grow with the level asymmetry. The growth of $S_2$ in particular is accompanied by the movement of H$_2$ towards H$_1$, and leads to their eventual collision and disappearance. The two segments $S_1$ and $S_2$ merge in this transition to give a single large segment of superradiant stability bounded by H$_3$, as shown in panel (d) of Fig. \ref{fig:asym_lm=0.6}.

\subsubsection*{Phase diagrams for small asymmetry}

\begin{figure}
    \centering
    \includegraphics[width=\columnwidth]{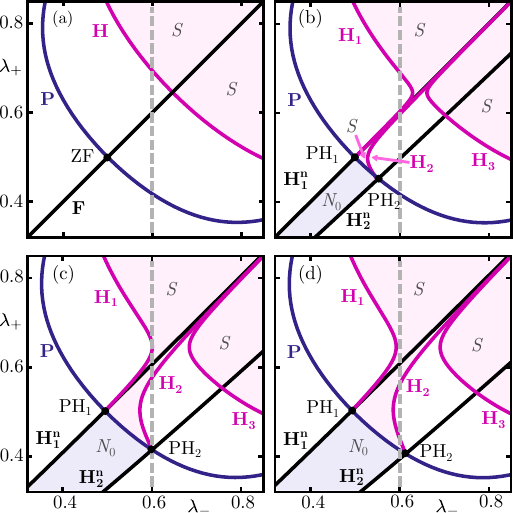}
    \caption{Phase diagrams for increasing asymmetry. The $(q,\omega_0)$ values used are those in Fig. \ref*{fig:asym_lm=0.6}, namely $(1,0)$ (a), $(0.999,0.001)$ (b), $(0.993,0.007)$ (c), $(0.99,0.01)$ (d). Curves of pitchfork, superradiant Hopf, and normal-phase Hopf bifurcations are labelled \textbf{P}, \textbf{H}, and \textbf{H}$^\textbf{n}$, respectively. Points of zero-flip and pitchfork-Hopf bifurcations are labelled ZF and PH. Regions of stable superradiance are labelled $S$ (pink shading), and regions of stable normal phase are lablled $N_0$ (blue shading). Dashed gray lines show the slices of the $(\lambda_-,\lambda_+)$-plane corresponding to the bifurcation diagrams in Fig. \ref*{fig:asym_lm=0.6}.}
    \label{fig:asym_prog1}
\end{figure}

The movement of the bifurcation points and stable segments in Fig. \ref{fig:asym_lm=0.6} is better understood by considering changes of the phase diagram in the $(\lambda_-,\lambda_+)$-plane. Figure ~\ref*{fig:asym_prog1} shows the phase diagrams in an appropriate region of parameter space for the values of $q$ and $\omega_0$ that we considered in Fig. \ref*{fig:asym_lm=0.6}. In Fig. ~\ref*{fig:asym_prog1}(a) we show the relevant part of the symmetric V configuration phase diagram, i.e., a part of Fig.~\ref*{fig:sym_diag} from the previous section; note that it does not include the origin, $(\lambda_-,\lambda_+)=(0,0)$. The region of stable superradiance is labelled $S$, which is delineated by the curve \textbf{H} of superradiant Hopf bifurcations. Curves of pitchfork and superradiant-flip bifurcations are also shown and labelled \textbf{P} and \textbf{F}, respectively; the latter is the diagonal, and it intersects \textbf{P} at a single zero-flip bifurcation, labelled ZF. The dashed vertical line at $\lambda_-=0.6$ corresponds to the one-parameter bifurcation diagram in Fig.~\ref*{fig:asym_lm=0.6}(a); its intersections with the curves \textbf{P}, \textbf{F}, and \textbf{H} occur at the locations of the corresponding codimension-one bifurcation points of the one-parameter bifurcation diagram.

Introducing level asymmetry breaks the balance of energy exchange that exists at the superradiant-flip bifurcation, and a topologically different phase diagram emerges. This can be seen in panels (b), (c), and (d) of Fig. \ref{fig:asym_prog1} which show three qualitatively equivalent phase diagrams, in the same region of the $(\lambda_-,\lambda_+)$-plane as panel (a), for increasing level asymmetry. The first effect of level asymmetry is the splitting of the supperradiant-flip curve \textbf{F} into two Hopf bifurcation curves \textbf{H}$^\textbf{n}_\textbf{1}$ and \textbf{H}$^\textbf{n}_\textbf{2}$ of the normal phase equilibrium N$_0$. These intersect \textbf{P} at the pitchfork-Hopf bifurcation points PH$_1$ and PH$_2$, where N$_0$ simultaneously has a pair of conjugate imaginary eigenvalues and a zero eigenvalue. Together with \textbf{P}, these curves enclose the region of normal phase stability $N_0$. The second effect of level asymmetry is the splitting of the superradiant Hopf bifurcation curve \textbf{H} into the curves \textbf{H}$_\textbf{1}$, \textbf{H}$_\textbf{2}$, and \textbf{H}$_\textbf{3}$. The part of \textbf{H} lying above \textbf{F} becomes the curve \textbf{H}$_\textbf{1}$, the part lying below \textbf{F} becomes the curve \textbf{H}$_\textbf{3}$, and between them is \textbf{H}$_\textbf{2}$. The curves \textbf{H}$_\textbf{1}$ and \textbf{H}$_\textbf{2}$ terminate at PH$_1$ and PH$_2$, respectively, when they each intersect \textbf{P}. They bound a region of superradiance together with \textbf{P}, while \textbf{H}$_\textbf{3}$ bounds another such region.

The differences between the diagrams in panels (b), (c), and (d) are only quantitative: the bifurcation curves mentioned above move as the asymmetry is changed, but their arrangement and intersections remain the same. As a consequence, the line $\lambda_-=0.6$ intersects the various bifurcation curves differently, producing some of the changes we observed in the one-parameter slices from Fig.~\ref*{fig:asym_lm=0.6}. For example, the line $\lambda_-=0.6$ intersects \textbf{H}$_\textbf{1}$ twice in panel (c), and not at all in panel (d). This corresponds to the collision and disappearance of the points H$_1$ and H$_2$ from Fig.~\ref*{fig:asym_lm=0.6}, which we can now see is not due to a qualitative change of the phase diagram. Overall, going from the largest asymmetry in panel (d) to no asymmetry in panel (a), these quantitative differences compound to recreate the phase diagram in the symmetric V configuration: \textbf{H}$^\textbf{n}_\textbf{1}$ and \textbf{H}$^\textbf{n}_\textbf{2}$ move closer together and towards the diagonal as the asymmetry decreases, \textbf{H}$_\textbf{1}$, \textbf{H}$_\textbf{2}$, and \textbf{H}$_\textbf{3}$ likewise move towards each other, and, consequently PH$_1$ and PH$_2$ do so as well. In the limit where $\omega_0\to0$ and $q\to1$, the curves \textbf{H}$^\textbf{n}_\textbf{1}$, \textbf{H}$^\textbf{n}_2$, and \textbf{H}$_\textbf{2}$ collapse onto the diagonal and become \textbf{F}, \textbf{H}$_\textbf{1}$ and \textbf{H}$_\textbf{3}$ join to become \textbf{H}, and PH$_1$ and PH$_2$ meet and become the point ZF.

\subsubsection*{Phase diagrams for large asymmetry}

\begin{figure}
    \centering
    \includegraphics[width=\columnwidth]{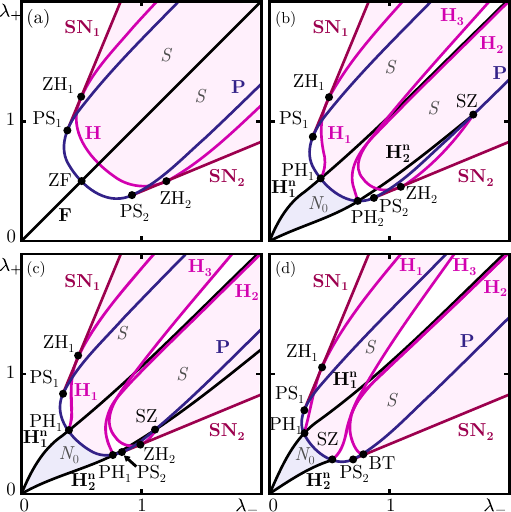}
    \caption{Phase diagrams for $(q,\omega_0)=(1,0)$ (a), $(0.9,0.1)$ (b), $(0.85,0.15)$ (c), and $(0.7,0.3)$ (d). Codimension-one curves and regions of equilibrium stability are coloured and labelled as before. Additional saddle-node curves are labelled \textbf{SN}. Codimension-two pitchfork-saddle-node (PS), pitchfork-Hopf (PH), symmetric double-zero (SZ), zero-Hopf (ZH), and Bogdanov-Takens (BT) bifurcation points are marked and labelled accordingly.}
    \label{fig:asym_prog2}
\end{figure}

Topological changes of the phase diagram in a larger region of the $(\lambda_-,\lambda_+)$-plane occur when the asymmetry is increased further, as shown in Fig. \ref*{fig:asym_prog2}. In panel (a) we once again show the phase diagram for the symmetric V configuration, but now over a larger range compared to Fig. \ref{fig:asym_prog1}(a) so that the two saddle-node bifurcation curves, \textbf{SN}$_\textbf{1}$ and \textbf{SN}$_\textbf{2}$, are also visible. These curves intersect \textbf{H} tangentially at two zero-Hopf bifurcation points ZH$_1$ and ZH$_2$, and they terminate at two pitchfork-saddle-node bifurcations PS$_1$ and PS$_2$. The zero-flip bifurcation is also shown and marked ZF.

As demonstrated in the previous section, introducing level asymmetry gives topologically different phase diagrams. Two such diagrams are shown in panel (b) of Fig. \ref*{fig:asym_prog2}, for $(q,\omega_0)=(0.9,0.1)$, and in panel (c) for $(q,\omega_0)=(0.85,0.15)$. As in Fig. \ref{fig:asym_prog1}, the three curves \textbf{H}$_\textbf{1}$, \textbf{H}$_\textbf{2}$, and \textbf{H}$_\textbf{3}$ of Hopf bifurcations appear. Two of these, \textbf{H}$_\textbf{1}$ and \textbf{H}$_\textbf{2}$, terminate at pitchfork-Hopf bifurcation points. In this larger range of the $(\lambda_-,\lambda_+)$-plane, we can now see that the third curve \textbf{H}$_\textbf{3}$ terminates at a symmetric double-zero bifurcation SZ, along with the normal-phase Hopf bifurcation curve \textbf{H}$^\textbf{n}_\textbf{2}$. As its name suggests, this bifurcation occurs when an equilibrium point has a zero eigenvalue of multiplicity two, and is subject to additional $\mathbb{Z}_2$ symmetry \cite*{carr}.

Increasing the level asymmetry from panel (b) to panel (c) of Fig. \ref{fig:asym_prog2} does not cause any qualitative changes, but the codimension-two bifurcation points PH$_2$, PS$_2$, ZH$_2$ and SZ move closer together. With a sufficiently large level asymmetry, these points meet and, consequently, the phase diagram changes topologically as shown in Fig. \ref{fig:asym_prog2}(d). Instead of changing criticality at the zero-Hopf point ZH$_2$ and terminating at SZ, the curve \textbf{H}$_\textbf{3}$ now ends at the Bogdanov-Takens bifurcation BT. In addition, \textbf{H}$_\textbf{2}$ no longer originates from the pitchfork-Hopf bifurcation point PH$_2$, but instead it ends at SZ. Further topological changes to the phase diagram do occur, but only when the level asymmetry is increased beyond the V configuration and into the L configuration \cite{adiv}, which is beyond the scope of this work.

\subsection{Bifurcations of periodic orbits}

Several of the codimension-two bifurcations described above imply the existence of certain bifurcations of periodic orbits nearby. For example, a pitchfork-Hopf bifurcation features pitchfork bifurcations of periodic orbits, and complicated periodic dynamics may arise in the vicinity of a zero-Hopf bifurcation \cite{kuznetsov}. Similarly to bifurcations of equilibria, the bifurcations of periodic orbits form the oscillatory phase diagram, which changes topologically with level asymmetry. To illustrate this, we focus in this section on two `case studies': small and moderate asymmetries. These show how some intricate transitions develop for oscillating solutions when increasing the level asymmetry, even within parameter regions where the equilibrium phase diagram does not change qualitatively.

\subsubsection*{Oscillations for small asymmetry}

\begin{figure}
    \centering
    \includegraphics[width=\columnwidth]{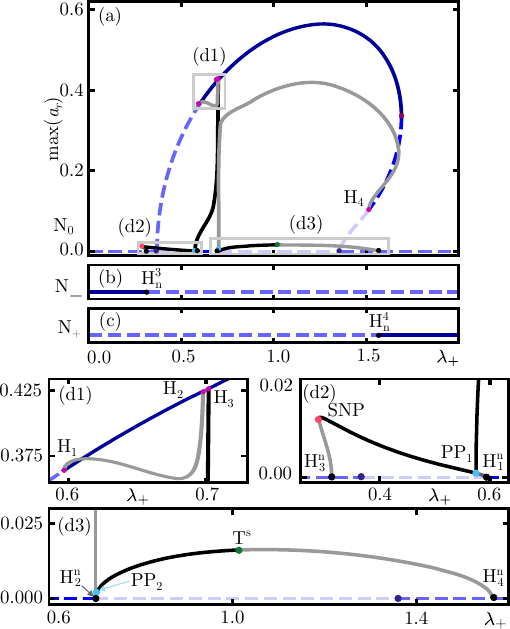}
    \caption{Bifurcation diagram for $\lambda_-=0.7$ and $(q,\omega_0)=(0.999,0.001)$. Panels (a), (b), and (c) show, respectively, N$_0$, N$_-$, N$_+$ and their bifurcating branches of equilibria and/or periodic orbits. Panels (d1), (d2), and (d3) are enlargements of the respective frames in panel (a). Bifurcations of equilibria are labelled as previously; relevant bifurcations of periodic orbits are also marked: saddle-node of periodic orbits SNP, pitchfork of periodic orbits PP, and torus T$^\text{s}$.}
    \label{fig:small_asym_lm=0.7}
\end{figure}

We begin with a slightly asymmetric level structure, corresponding to $(q,\omega_0)=(0.999,0.001)$, where new periodic orbit branches emerge and stable oscillations develop. Figure \ref*{fig:small_asym_lm=0.7} shows the one-parameter bifurcation diagram for $\lambda_-=0.7$, where we now separate the three normal phase equilibria, N$_0$, N$_-$, and N$_+$. These correspond, respectively, to states where all the atoms are in one of the $m=0$, $-1$, and $+1$ sublevels, and may undergo pitchfork or Hopf bifurcations leading to equilibrium and oscillatory superradiance, respectively.

Here N$_0$ is the primary bifurcating equilibrium point, as shown in panel (a), where it undergoes a pitchfork bifurcation leading to equilibrium superradiance. Along the superradiant branch, equilibria undergo several Hopf bifurcations labelled H$_{1-4}$. Two of these, H$_2$ and H$_3$, split from the superradiant-flip bifurcation; compare with Fig.~\ref*{fig:sym_v_one_par}(a) showing the bifurcation diagram for the same $\lambda_-$ in the symmetric V configuration.

As in Fig.~\ref*{fig:sym_v_one_par}(a), H$_1$ is subcritical and gives rise to a branch of unstable periodic orbits; see Fig. \ref*{fig:small_asym_lm=0.7}(d1). Instead of connecting to H$_4$, as it does in the symmetric case, this branch ends at H$_2$, which is one of the bifurcations that split from the superradiant-flip bifurcation. The other, H$_3$, is supercritical and gives rise to a nearly vertical branch of stable, asymmetric periodic orbits.

This branch connects H$_3$ to a pitchfork bifurcation PP$_1$ of a symmetric periodic orbit, shown in panel (d2). The symmetric orbits along this branch have a small amplitude in $a_r$; this is consistent with their gradual appearance as the asymmetry increases and the transitions from the $m=0$ sublevel to the $m=+1$ and $m=-1$ sublevels become increasingly imbalanced. In the direction of increasing $\lambda_+$, the branch develops from a saddle-node bifurcation SNP of periodic orbits, whose unstable orbit vanishes in a subcritical Hopf bifurcation H$^n_3$ of N$_-$, labelled and shown in Fig. \ref*{fig:small_asym_lm=0.7}(b) as well. The stable periodic orbit that emerges from the point SNP loses stability in the aforementioned pitchfork bifurcation PP$_1$, and vanishes at a Hopf bifucation, H$^n_1$, of N$_0$.

A second Hopf bifurcation H$^n_2$ of N$_0$ leads to the creation of a second symmetric periodic orbit, as seen in Fig. \ref*{fig:small_asym_lm=0.7}(d3). Following this bifurcation, the periodic orbit is unstable, although it gains stability shortly thereafter in a subcritical pitchfork bifurcation PP$_2$. The unstable asymmetric orbit branch that emerges from PP$_2$ is also nearly vertical, and subsequently ends at H$_4$. The appearance of these nearly vertical branches is no coincidence, given they emerge (whether directly or via bifurcations of another periodic orbit) from Hopf bifurcations that split from the superradiant-flip bifurcation. Recall that at the superradiant-flip bifurcation a surface foliated by infinitely many orbits connects the superradiant equilibria, and, in a bifurcation diagram, this would be represented by a vertical branch of periodic orbits. These periodic orbits `extend' over a (small) range of parameters once level asymmetry is introduced and the energy balance at the superradiant-flip point is broken.

\begin{figure}
    \centering
    \includegraphics[width=\columnwidth]{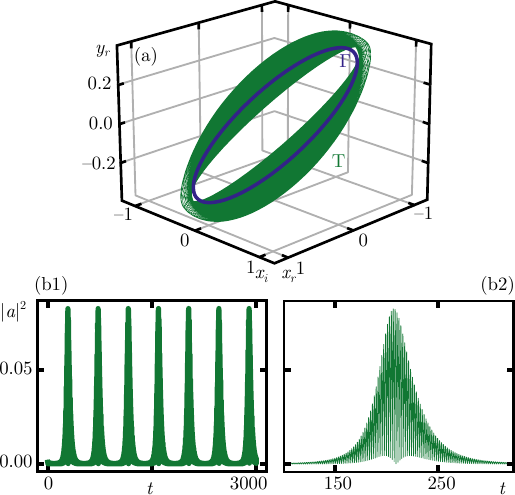}
    \caption{Segment of a trajectory on a torus T for $\lambda_-=0.7$, $\lambda_+=1.1$, and $(q,\omega_0)=(0.999,0.001)$. Panel (a) shows a projection of the trajectory and the periodic $\Gamma$ from which it came. Panel (b1) shows a time series of the cavity mode population $|a|^2$, while panel (b2) shows an individual pulse from panel (b1).}
    \label{fig:small_asym_tor}
\end{figure}

After gaining stability at the point PP$_2$, the symmetric periodic orbit branch loses stability again at a supercritical torus bifurcation T$^\text{s}$ and then disappears at a Hopf bifurcation H$^n_4$ of N$_+$, which is also shown in Fig. \ref*{fig:small_asym_lm=0.7}(c). The torus bifurcation leads to the creation of stable \textit{two-frequency} trajectories which lie on the surface of a symmetric torus. Figure \ref*{fig:small_asym_tor}(a) shows how a segment of one such trajectory effectively covers a torus T in phase space near the now unstable periodic orbit $\Gamma$ from which it developed. It follows that (sufficiently close to the bifurcation) oscillations of trajectories on T `around' the middle hole of the torus have approximately the same frequency as $\Gamma$. The difference between these trajectories and $\Gamma$ is that they have a second direction of oscillation `through' the hole of the torus, with a separate frequency.

This is also apparent in the time series of the cavity population along the trajectory segment from Fig. \ref*{fig:small_asym_tor}(a), which is shown in Fig. \ref*{fig:small_asym_tor}(b1) and (b2). The two-frequency dynamics manifests itself as fast oscillations that are subject to deep modulation at a much lower frequency. At the shown time scale of Fig. \ref{fig:small_asym_tor}(b1), the latter give the appearance of individual pulses. The enlargement in Fig. \ref{fig:small_asym_tor}(b2) shows that $|a|^2$ actually oscillates (in a pulsing fashion) at a much faster rate; the ratio of the two frequencies is approximately 1:86.

Numerical continuation of tori is possible \cite*{tori}, but very challenging and beyond the scope of this work. For our purposes it is sufficient to note that, through careful numerical integration, the torus T was found to vanish in a bifurcation of the normal phase tori. This appears to happen before the saddle-node bifurcation of the superradiant equilibria for all choices of $\lambda_-$ we considered, and the torus appears to be stable throughout.

\begin{figure}
    \centering
    \includegraphics[width=\columnwidth]{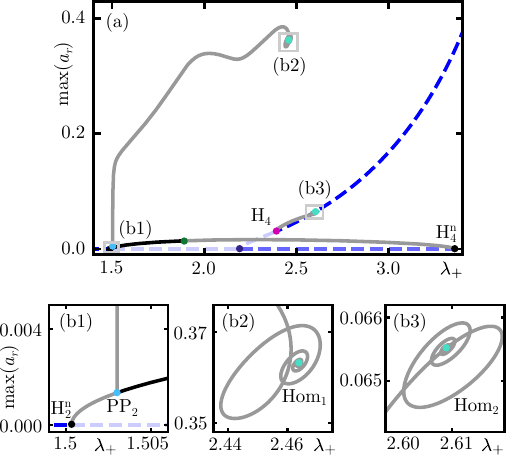}
    \caption{Bifurcations of periodic orbits for $\lambda_-=1.5$ and $(q,\omega_0)=(0.999,0.001)$. Hopf and pitchfork bifurcations of periodic orbits are labelled as before. Homoclinic bifurcations are labelled Hom. Panels (b1), (b2), and (b3) are magnifications of the diagram shown in panel (a) near PP$_2$, Hom$_1$, and Hom$_2$.}
    \label{fig:small_asym_lm=1.5_homs}
\end{figure}

For larger $\lambda_-$ the bifurcations of periodic orbits change little from Fig.~\ref*{fig:small_asym_lm=0.7}. As in the symmetric V configuration, the main difference is the disappearance of two periodic orbits via a pair of homoclinic bifurcations; refer to Fig.~\ref{fig:sym_v_one_par}(b). This is shown for $\lambda_-=1.5$ in Fig.~\ref*{fig:small_asym_lm=1.5_homs}. In panel (a) we see that there no longer exists a single branch of periodic orbits that connects PP$_2$ and H$_4$. Instead, one branch emerges from PP$_2$ as seen in panel (b1), and ends at a homoclinic bifurcation labelled Hom$_1$, shown in panel (b2). A second periodic orbit branch emerges from H$_4$ and ends at a second homoclinic bifurcation labelled Hom$_2$, which is shown in panel (b3).

\begin{figure*}
    \centering
    \includegraphics{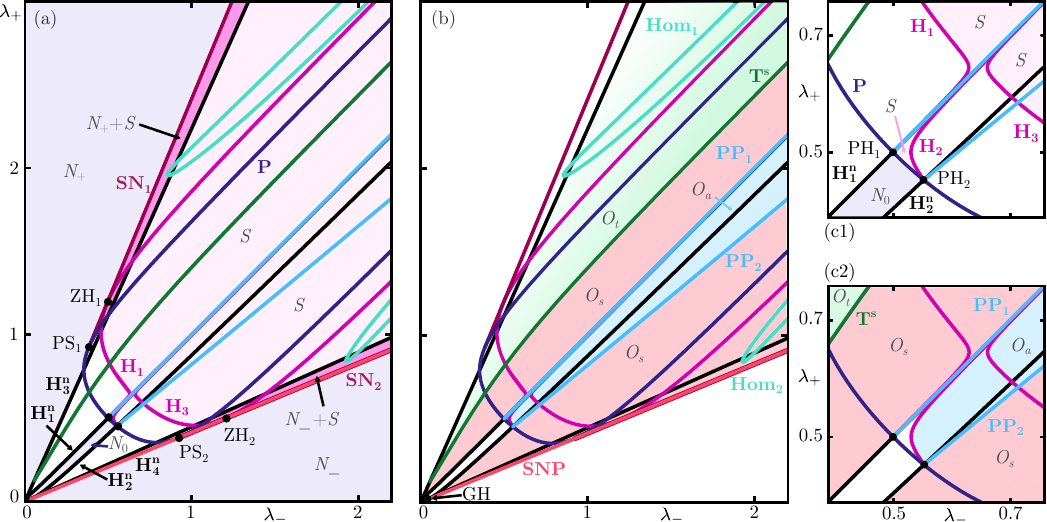}
    \caption{Equilibrium (a) and oscillatory (b) phase diagrams for the asymmetric V configuration with $(q,\omega_0)=(0.999,0.001)$. Regions shaded in the same colour denote the same phase. These are the normal phases $N_0$, $N_-$, and $N_+$, superradiance $S$, symmetric oscillations $O_\text{s}$, asymmetric oscillations $O_\text{a}$, and oscillations on a torus $O_\text{t}$. Regions of equilibrium bistability are labelled $N_\pm+S$ (depending on the normal-phase equilibrium involved). Panels (c1) and (c2) are enlargements of the parameter regions in (a) and (b).}
    \label{fig:small_asym_diag}
\end{figure*}

Following the above bifurcations of periodic orbits as we vary $\lambda_-$ allows us to construct the phase diagrams in Fig. \ref*{fig:small_asym_diag}. In panels (a) and (b) we show the bifurcation curves and regions of stability over a large portion of the $(\lambda_-,\lambda_+)$-plane, while panels (c1) and (c2) are enlarged regions of panels (a) and (b). Bifurcations of equilibria are labelled as before, and there are several additional curves denoting bifurcations of periodic orbits.

Two of these, \textbf{PP}$_\textbf{1}$ and \textbf{PP}$_\textbf{2}$, complete the bifurcation scenarios at the pitchfork-Hopf bifurcations, PH$_1$ and PH$_2$. This may be seen more clearly in panel (c1), where \textbf{PP}$_\textbf{1}$ and \textbf{PP}$_\textbf{2}$ emerge from PH$_1$ and PH$_2$ respectively, alongside the Hopf bifurcation curves \textbf{H}$_\textbf{1}$ and \textbf{H}$_\textbf{2}$.

The curve \textbf{SNP} closely follows the curve \textbf{SN}$_\textbf{2}$, meaning that, for large enough $\lambda_-$, the onset of superradiance is marked by the simultaneous appearance of an equilibrium point and a periodic orbit. Closer inspection reveals that \textbf{SNP} emerges from a generalized-Hopf bifurcation point GH along the curve \textbf{H}$^\textbf{n}_\textbf{2}$ of Hopf bifurcations of N$_0$. It occurs (comparatively) close to $(\lambda_-,\lambda_+)=(0,0)$. This suggests that the point GH approaches $(\lambda_-,\lambda_+)=(0,0)$ in the limit where $\omega_0\to0$ and $q\to1$ (the symmetric V configuration), which is consistent with the gradual appearance of the aforementioned symmetric periodic orbits as level asymmetry is increased.

Altogether, these bifurcation curves enclose regions of stability for the various phases. Figs.~\ref{fig:small_asym_diag}(a) and (c1) show the equilibrium phase diagram, where there is little change from the symmetric configuration. Superradiant regions are marked $S$, while the various normal-phase regions are marked $N_0$, $N_+$, and $N_-$. Similar to the two-level Dicke model, there are narrow regions of equilibrium bistability, labelled $N_\pm+S$ in panel (a), which are bounded by curves of normal-phase Hopf and saddle-node bifurcations.

Figures \ref{fig:small_asym_diag}(b) and (c2) show the associated oscillatory phase diagram, which features more dramatic differences from the symmetric V configuration, namely the appearance of large oscillatory regions. We separate these into symmetric, asymmetric, and two-frequency oscillations, respectively labelled $O_s$, $O_a$, and $O_t$, which altogether extend across large and small values of $\lambda_+$ and $\lambda_-$. Note that we make the distinction between symmetric and asymmetric periodic orbits. The former connect two regions of phase-space, while the latter do not, and this affects the cavity population and, therefore, experimental observations.

Comparing the phase diagrams in Figs.~\ref{fig:small_asym_diag} (a) and (b) one notices considerable overlap between regions of superradiance and the various regions of oscillation. For example, superradiant equilibria may coexist with normal phase equilibria as well as symmetric periodic orbits in the region labelled $N_-+S$. This, as well as other regions of multistability, may prove interesting in the quantum-mechanical regime where quantum fluctuations cannot be ignored. In analogy with the two-level Dicke model, quantum fluctuations have been shown to drive switching between bistable phases, and allow the formation of a hysteresis loop \cite*{stitely_bistab}. In our case, normal phase tori may also be stable in any of these regions, thereby enriching the dynamics by making complex switching patterns possible. Hence, the intricacies of these behaviours in a semiclassical description are critical for understanding the corresponding quantum-mechanical behaviour.

\subsubsection*{Oscillations for Moderate asymmetry}

We now consider a moderate asymmetry corresponding to $(q,\omega_0)=(0.95,0.05)$, with the remaining parameters fixed as before at $\kappa=\omega=1$. Note that the equilibrium phase diagram for this choice of $q$ and $\omega_0$ is qualitatively equivalent to that in the previous section. However, there are qualitative changes to the oscillatory phase diagram that are the result of passing through codimension-three bifurcations of periodic orbits.

\begin{figure}
    \centering
    \includegraphics[width=\columnwidth]{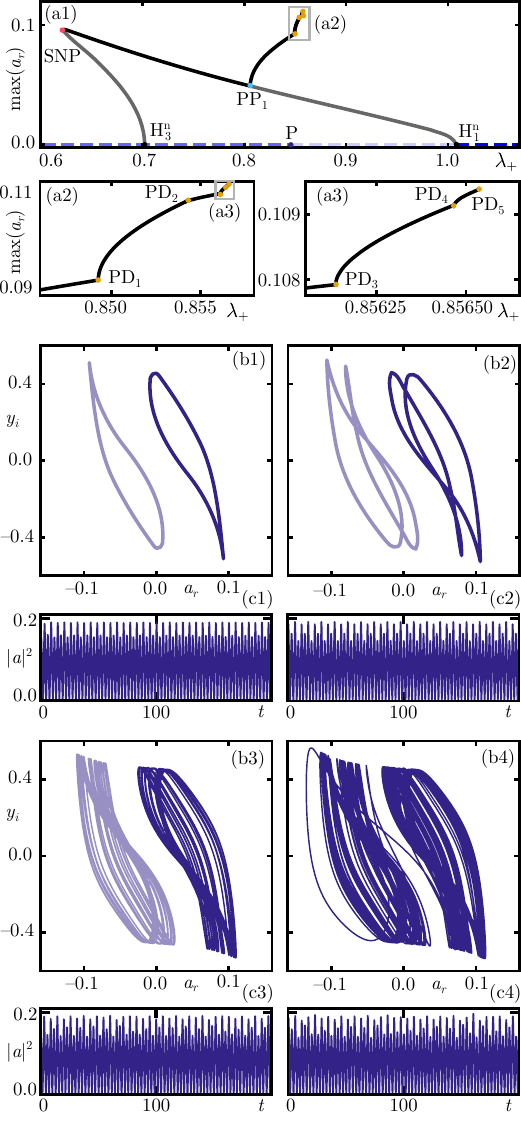}
    \caption{Bifurcation diagram of a symmetric periodic orbit branch for $\lambda_-=1.5$ and $(q,\omega_0)=(0.95,0.05)$, and its path to chaos in the $(a_r,y_i)$-projection. Panel (a1) shows the bifurcation diagram. Panels (a2) and (a3) show successive magnifications of the diagram in panel (a1), showing in total five period-doubling bifurcations labelled PD. Panels (b1)--(b3) show a period-doubling cascade leading to the formation of two asymmetric chaotic attractors, while panel (b4) shows a symmetric chaotic attractor resulting from the collision of the attractors in (b3). The corresponding values of $\lambda_+$ are $0.8492$ (b1), $0.8541$ (b2), $0.857$ (b3), and $0.859$ (b4). Panels (c1)--(c4) show the cavity population along the respective trajectories from panels (b1)--(b4).}
    \label{fig:mod_asym_cascade}
\end{figure}

The first major change involves the symmetric periodic orbit branch originating from the Hopf bifurcation H$^n_3$ of N$_+$. Figure \ref*{fig:mod_asym_cascade}(a1) shows the one-parameter bifurcation diagram of this branch. First, a pair of periodic orbits appears through a saddle-node bifurcation, SNP. The unstable periodic orbit disappears at H$^n_3$, while the stable one eventually loses stability in a pitchfork bifurcation PP$_1$. After doing so it vanishes in a Hopf bifurcation H$^n_1$ of N$_0$. So far, this is exactly the same bifurcation scenario that we encountered in the previous section; see Fig.~\ref*{fig:small_asym_lm=0.7}(d2). 

The difference lies in the branches of asymmetric periodic orbits that emerge from PP$_1$: where they previously underwent no further transitions and disappeared in superradiant Hopf bifurcations, they now undergo a period-doubling cascade as shown in Figs.~\ref{fig:mod_asym_cascade}(a2) and (a3). In a supercritical period-doubling bifurcation, such as the shown points PD$_{1-5}$, a periodic orbit loses stability and another stable orbit with twice the period appears. Cascades of infinitely many successive period-doubling bifurcations, within a finite parameter range, are a common route to chaotic dynamics \cite*{guck}. Each successive period-doubling happens sooner than its predecessor and, in the limit, a chaotic attractor is formed.

Figures~\ref*{fig:mod_asym_cascade}(b1)--(b3) illustrate this route to chaos in phase space, and Figs.~\ref*{fig:mod_asym_cascade}(c1)--(c3) show the corresponding cavity population. In panel (b1) there are two asymmetric period-1 orbits, related by $\mathbb{Z}_2$ symmetry, which emerged from the pitchfork bifurcation PP$_1$, with a regularly oscillating cavity population in panel (c1). After the (simultaneous) period-doubling bifurcation PD$_1$ of these period-1 orbits, two attracting period-2 orbits are present; see panel (b2) and the corresponding regular oscillations of the cavity population, in panel (c2), at approximately twice the period. After the cascade two separate, symmetry-related chaotic attractors remain, as shown in panel (b3), for which the cavity population in panel (c3) oscillates chaotically.

As $\lambda_+$ is increased the two chaotic attractors approach each other and then merge when they reach the symmetry subspace of $\mathcal{T}_{\mathbb{Z}_2}$ from Eq. (\ref{eq:z2_sym}). This creates a large, symmetric attractor, which is shown in panel (b4) with the corresponding cavity population in panel (c4). This transition marks a significant change, since the symmetric attractor bridges two previously disconnected regions of phase-space. Similar merging phenomena of chaotic attractors in $\mathbb{R}^4$ have been closely related to, and facilitated by, homoclinic orbits \cite{giraldo2022,bitha2023}. In this context, the homoclinic orbits additionally determine the observed switching dynamics; that is, how trajectories move between different parts of phase space. Analysis of switching dynamics in this fashion is made possible by numerical continuation of the relevant homoclinic orbits. This requires the use of sophisticated techniques, such as Lin's method \cite{krauskopf2008}, which ultimately allow one to paint a detailed dynamical picture of such chaotic dynamics within large parameter regions.

\begin{figure}[t!]
    \centering
    \includegraphics[width=\columnwidth]{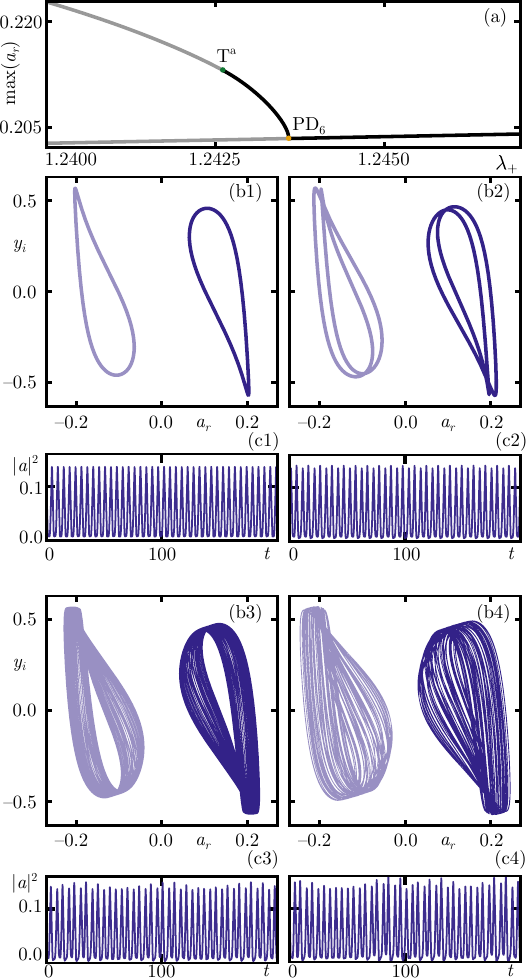}
    \caption{Disappearance of a pair of asymmetric chaotic attractors for $\lambda_-=1.5$ and $(q,\omega_0)=(0.95,0.05)$. Panel (a) shows the bifurcation diagram nearby, with marked torus bifurcations T$^\text{a}$ and period-doubling bifurcation PD$_6$. Panels (b1)--(b4) show trajectories for $\lambda_+=1.24$ (b1), $1.242$ (b2), $1.2427$ (b3), and $1.244$ (b4). More specifically, (b1) shows a pair of chaotic attractors, (b2) shows segments of trajectories on a pair of asymmetric tori, (b3) shows a pair of period-2 orbits, and (b4) shows a pair of period-1 orbits. Panels (c1)--(c4) show the cavity population along the respective trajectories from panels (b1)--(b4).}
    \label{fig:mod_asym_reverse}
\end{figure}

However, we focus here on the boundaries of chaotic behaviour and not in their intricate details. Hence, it is sufficient for us to present the bifurcations that mark the end of chaotic dynamics, or, conversely, the route to those chaotic dynamics in the direction of decreasing $\lambda_+$. This is illustrated in Fig. \ref{fig:mod_asym_reverse}, where panel (a) shows the one-parameter bifurcation diagram of a branch of periodic orbits. For large $\lambda_+$, there exist a pair of stable, asymmetric period-1 orbits. Their projection onto the $(a_r,y_i)$-plane in panel (b1) and the corresponding cavity population is shown in panel (c1). This pair undergoes simultaneous period-doubling bifurcations, labelled PD$_6$ in Fig. \ref{fig:mod_asym_reverse}(a), thereby losing stability. This results in the creation of two stable period-2 orbits, whose phase-space projection and cavity population are shown in Figs. \ref{fig:mod_asym_reverse}(b2) and (c2). These period-2 orbits lose stability in the supercritical torus bifurcation T$^\text{a}$, leading to stable two-frequency oscillations on the surfaces of a pair of tori, as shown in Figs. \ref{fig:mod_asym_reverse}(b3) and (c3). These tori lose smoothness in a process known as \textit{torus breakdown} \cite{sano1983,krauskopf2000}, suddenly creating a pair of asymmetric chaotic attractors as in panels (b4) and (c4). These attractors eventually merge, as outlined above, to form the large symmetric attractor from Fig. \ref{fig:mod_asym_cascade}.

\begin{figure*}
    \centering
    \includegraphics{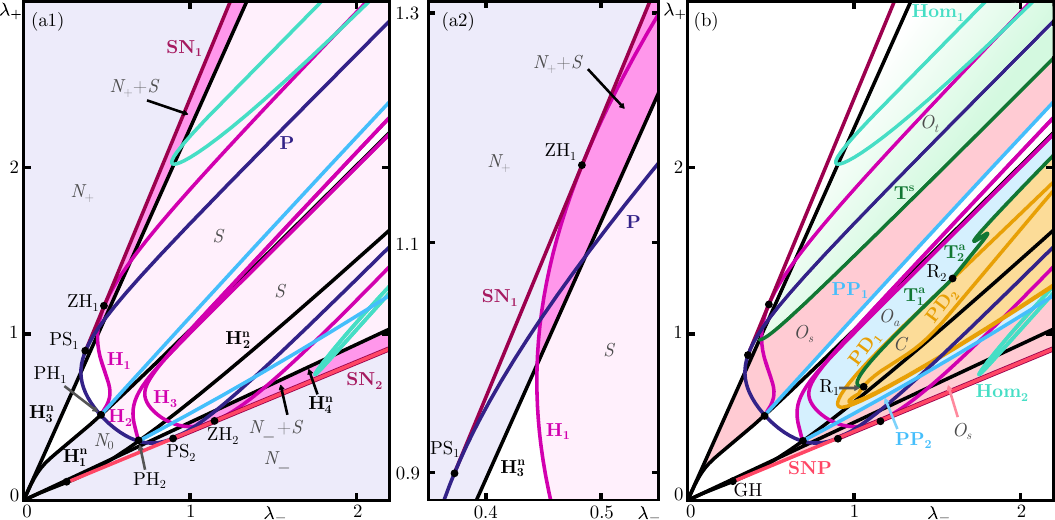}
    \caption{Phase diagrams for the moderately asymmetric V configuration, with $(q,\omega_0)=(0.95,0.05)$. Panels (a1) and (a2) show the equilibrium phase diagram, while panel (b) shows the oscillatory one. Regions of stability are shaded as in Fig. \ref*{fig:small_asym_diag}, with an additional region of complicated dynamics labelled $C$ and shaded in yellow.}
    \label{fig:mod_asym_diag}
\end{figure*}

Fig.~\ref{fig:mod_asym_diag} summarizes our findings by showing which type of dynamics can be found where in the $(\lambda_-,\lambda_+)$-plane for the represantative choice $(q,\omega_0)=(0.95,0.05)$ of a moderately asymmetric V configuration. To aid the interpretation of the overall structure, we show separate phase diagrams for the equilibria in panel~(a1) of Fig. \ref{fig:mod_asym_diag}, and for the oscillatory solutions in its panel~(b); the observable dynamics for any point in the $(\lambda_-,\lambda_+)$-plane is, hence, given by the respective stable objects detailed in each of these two diagrams. A characteristic feature of the overall phase diagram is a high degree of coexistence between different types of solutions in certain parameter regions. In particular, Fig.~\ref{fig:mod_asym_diag} highlights the regions where one may find chaotic or other complicated dynamics, and illustrates how this kind of behavior can be reached via the routes to chaos from Figs. \ref{fig:mod_asym_cascade} and \ref{fig:mod_asym_reverse}. 

The equilibrium phase diagram Fig. \ref{fig:mod_asym_diag}(a) is qualitatively the same as that for $(q,\omega_0)=(0.999,0.001)$ in Fig.~\ref{fig:small_asym_diag}(a). This is to say, that both have the same curves of equilibrium bifurcations, with the same arrangement in terms of their intersections and, hence, the equivalent regions where one finds normal-phase and/or superradiant equilibria. However, for the more asymmetric case in Fig.~\ref{fig:mod_asym_diag}(a) the curves have moved further apart, so that the overall structure of the equilibrium phase diagram is easier to discern at the shown scale. In particular, as is further illustrated by the enlargemt panel~(a2), the curve of saddle-node bifurcations \textbf{SN}$_\textbf{1}$ is intersected tangentially by the curve \textbf{P} of pitchfork bifurcations at the pitchfork-saddle-node point PS$_\text{1}$, and by the curve \textbf{H}$_\textbf{1}$ of Hopf bifurcations at the zero-Hopf point ZH$_\text{1}$. The remaining bifurcation curves intersect similarly at additional pitchfork-saddle-node and zero-Hopf points, as well as pitchfork-Hopf and generalised Hopf points as shown. Each such point of codimension-two bifurcation marks a change in criticality along the relevant bifurcation curves and, thus, a change in which curves bound the regions of normal-phase stability, superradiant stability, and stability of both.

The oscillatory phase diagram in Fig. \ref{fig:mod_asym_diag}(b) is effectively a refinement of that in Fig. \ref{fig:small_asym_diag}(b) for the small asymmetry of $(q,\omega_0)=(0.999,0.001)$. Again, the previously shown and discussed curves of saddle-node, pitchfork, torus, saddle-node, and homoclinic bifurcations of periodic solutions have moved further apart for $(q,\omega_0)=(0.95,0.05)$, so that they are easier to distinguish in Fig. \ref{fig:mod_asym_diag}(b), but they have the same overall arrangement. Notice, for example, how the two pitchfork bifurcation curves \textbf{PP}$_\textbf{1}$ and \textbf{PP}$_\textbf{2}$ originate from the pitchfork-Hopf points PH$_\text{1}$ and PH$_\text{2}$, respectively, while the curve \textbf{SNP} originates from the generalised-Hopf point GH. These curves divide the $(\lambda_-,\lambda_+)$-plane into the same regions $O_s$, $O_a$, and $O_t$ of symmetric, asymmetric, and two-frequency oscillations, which have significant overlap with the regions of equilibrium stability shown in Fig. \ref{fig:mod_asym_diag}(a1). Hence, for certain choices of $\lambda_+$ and $\lambda_-$, one finds two or more stable dynamical objects of both stationary and oscillatory nature. For example, the region labelled $N_-+S$ in panel~(a1) overlaps with the region labelled $O_s$ in panel~(b); for any $\lambda_+$ and $\lambda_-$ in this area of overlap, the phase space contains a stable normal-phase equilibrium, a pair of stable superradiant equilibria, and a stable (symmetric) periodic orbit.

As an additional element of the phase diagram, Fig. \ref{fig:mod_asym_diag}(b) also shows a region labeled C, which is bounded by the curves \textbf{PD}$_\textbf{1}$ and \textbf{PD}$_\textbf{2}$ of period-doubling bifurcations, and the curves \textbf{T}$^\textbf{a}_\textbf{1}$ and \textbf{T}$^\textbf{a}_\textbf{2}$ of torus bifurcations.
These are the bifurcations we identified as part of the transitions to chaotic dynamics in Figs. \ref{fig:mod_asym_cascade} and \ref{fig:mod_asym_reverse}, and the corresponding  curves provide a good indication of where in the $(\lambda_-,\lambda_+)$-plane complicated and chaotic dynamics can be found. The curves \textbf{PD}$_\textbf{1}$ and \textbf{T}$^\textbf{a}_\textbf{1}$, and \textbf{PD}$_\textbf{2}$ and \textbf{T}$^\textbf{a}_\textbf{2}$, respectively, meet at codimension-two 1:2 resonance points R$_1$ and R$_2$; here the bifurcating periodic orbit has double $-1$ Floquet multipliers and the criticality of the respective two curves changes \cite{kuznetsov}. The region C is bounded by the respective supercritical parts of these curves, and it lies entirely in the region S of stable equilibrium superradiance; compare panels~(a1) and~(b) of Fig. \ref{fig:mod_asym_diag}. This additional region is significant not only because it involves chaotic dynamics, but also because it arises from only a slight change in $q$ and $\omega_0$, in a regime of quite small asymmetry of the V configuration. Increasing the asymmetry further may well involve further and possibly dramatic changes to the oscillatory phase diagram; however, its further investigation remains an interesting subject beyond the scope of this paper. 

%###########################################
%                Conclusion
%###########################################

\section{Conclusion}

We studied the nonlinear semiclassical dynamics of the spin-1 Dicke model with a bifurcation-oriented approach. By making use of both linear and quadratic Zeeman shifts, we were able to consider a general model where the atomic energy level structure is variable. We focused on the V configuration, where the $m=0$ magnetic sublevel has the lowest energy.

We first considered the symmetric V configuration, where the dynamics are comparatively simple and equilibrium superradiance emerges either through saddle-node or Hopf bifurcations. We also detailed the superradiant-flip bifurcation, which is brought about by an energy balance between atomic transitions and cavity decay. As we showed, the superradiant-flip breaks down with the introduction of asymmetry to the energy level structure. Several behaviours emerge as a result: both periodic and two-frequency stable oscillations, multistability between different types of behaviour, and chaotic dynamics. These were presented in detailed phase diagrams for various degrees of level asymmetry.

Our work can be taken in several future directions. First, intricate transitions within regions of chaotic or complicated dynamics can be analysed in more detail. For example, more sophisticated methods for the continuation of homoclinic orbits can be used, such as Lin's method \cite*{krauskopf2008}. Second, the implications of these semiclassical behaviours for the fully quantum-mechanical regime remain largely unexplored. They could be studied in the spirit of previous work on similar models \cite*{stitely_bistab,stitely2023}, through the use of quantum trajectory theory, or by considering the evolution of higher-order operator expectations.

\begin{acknowledgments}
The contribution of S.P. to this work was supported in part by grant NSF PHY-2309135 to the Kavli Institute for Theoretical Physics (KITP). The authors would like to thank Kevin Stitely for his helpful discussions and insights.
\end{acknowledgments}

%###########################################
%                Appendix
%###########################################

\appendix

\section{Stereographic projection with respect to N$_-$}\label{apdx:proj}

Following the procedure in Sec. \ref{sec:model}, we make a phase rotation and take a stereographic projection of Eqs. (\ref{eq:eoms1})--(\ref{eq:eoms2}). Here we do so with respect to N$_-$. Defining $r_-=|b_-|$ and $\phi_-=\arg(b_-)$ we make the transformation
\begin{equation}
    (b_+,b_-,b_0)\to(u,v)=\left(\frac{b_+e^{-i\phi_-}}{1-r_-},\frac{b_0e^{-i\phi_-}}{1-r_-}\right)\,.
\end{equation}

The equations of motion in this coordinate system are
\begin{widetext}
\begin{align}
    \dot{a}&=-(\kappa+i\omega)a-\frac{2i(|u|^2+|v|^2-1)}{(|u|^2+|v|^2+1)^2}(\lambda_-v+\lambda_+v^*)-\frac{4i(\lambda_+u^*v+\lambda_-uv^*)}{(|u|^2+|v|^2+1)^2}\,,\label{eq:sys5}\\
    \dot{u}&=-2i\omega_0u-iAv+\frac{iu}{|u|^2+|v|^2-1}(A^*v+Av^*)-\frac{iu}{2}(A^*v-Av^*)\,,\\
    \dot{v}&=i(q-\omega_0)v-iA^*u-\frac{iA}{2}(|u|^2+|v|^2-1)+\frac{iv}{|u|^2+|v|^2-1}(A^*v+Av^*)-\frac{iv}{2}(A^*v-Av^*)\,.
\end{align}
\end{widetext}
\noindent where $A=\lambda_-a+\lambda_+a^*$.

\section{Location of superradiant equilibria in the symmetric V configuration}\label{apdx:eqs}

In the symmetric V configuration, when $\omega_0=0$, Eqs. (\ref{eq:dbp}) and (\ref{eq:dbm}) become
\begin{align}
    \dot{b}_+&=-iqb_+-i(\lambda_-a+\lambda_+a^*)b_0\,,\\
    \dot{b}_-&=-iqb_--i(\lambda_+a+\lambda_-a^*)b_0\,.
\end{align}
\noindent Equilibria of these equations must also correspond to equilibria of the system given by Eqs. (\ref{eq:sys3})--(\ref{eq:sys4}), modulo their phase invariance. Hence, we look for $b_+$ and $b_-$ that satisfy
\begin{align}
    qb_++(\lambda_-a+\lambda_+a^*)b_0&=0\,,\label{eq:eqs_1}\\
    qb_-+(\lambda_+a+\lambda_-a^*)b_0&=0\,.\label{eq:eqs_2}
\end{align}

Next, as in Section \ref{sec:proj}, we introduce $r_0=|b_0|$ and $\phi_0=\arg(b_0)$. Multiplying Eqs. (\ref{eq:eqs_1}) and (\ref{eq:eqs_2}) by $e^{-i\phi_0}/(1-r_0)$, and using the definitions of $x$ and $y$, gives
\begin{align}
    qx+(\lambda_-a+\lambda_+a^*)\frac{r_0}{1-r_0}&=0\,,\label{eq:eqs_3}\\
    qy+(\lambda_+a+\lambda_-a^*)\frac{r_0}{1-r_0}&=0\,.\label{eq:eqs_4}
\end{align}
\noindent After taking the difference between Eq. (\ref{eq:eqs_3}) and the conjugate of Eq. (\ref{eq:eqs_4}) and simplifying, one gets $x=y^*\iff y=x^*$. Hence, any equilibria in the symmetric V configuration must lie on the hyperplane $x=y^*$.

In the case where $\lambda_+=\lambda_-=\lambda$, Eqs. (\ref{eq:eqs_3}) and (\ref{eq:eqs_4}) become
\begin{align}
    qx+2\lambda a_r\frac{r_0}{1-r_0}&=0\,,\\
    qy+2\lambda a_r\frac{r_0}{1-r_0}&=0\,,
\end{align}
\noindent from which it is easy to see that the equilibria must also lie on the hyperplane $x=y$, with $x_i=y_i=0$, i.e., $x$ and $y$ must be real.

%###########################################
%                References
%###########################################
\bibliography{refs.bib}\label{sec:refs}

\end{document}